\documentclass[sigconf]{acmart}
\usepackage{amsmath,amsfonts}
\usepackage{algorithm,algorithmicx,algpseudocode}
\usepackage{graphicx}
\usepackage{textcomp}
\usepackage{float}
\usepackage{multirow}
\usepackage{tabularx}  
\usepackage{enumitem}
\usepackage{adjustbox}
\usepackage{siunitx}
\usepackage{booktabs}
\usepackage{array}

\usepackage{balance}

\setcopyright{none}
\renewcommand\footnotetextcopyrightpermission[1]{}
\AtBeginDocument{%
  }

\begin{document}

\title{VRSafe: A Secure Virtual Keyboard to Mitigate Keystroke Inference in Virtual Reality}

\titlenote{Accepted at ACM CODASPY 2026}

\author{Yijun Yuan}

\orcid{0009-0008-0751-9207}
\affiliation{%
  \institution{University of Pittsburgh}
  \city{Pittsburgh}
  \state{PA}
  \country{USA}
}
\email{yiy95@pitt.edu}

\author{Na Du}

\affiliation{
  \institution{University of Pittsburgh}
  \city{Pittsburgh}
  \state{PA}
  \country{USA}
}
\email{na.du@pitt.edu}

\author{Adam J. Lee}

\affiliation{
  \institution{University of Pittsburgh}
  \city{Pittsburgh}
  \state{PA}
  \country{USA}
}
\email{adamlee@pitt.edu}

\author{Balaji Palanisamy}

\affiliation{
  \institution{University of Pittsburgh}
  \city{Pittsburgh}
  \state{PA}
  \country{USA}
}
\email{bpalan@pitt.edu}

\renewcommand{\shortauthors}{Yijun Yuan, Na Du, Adam J. Lee, and Balaji Palanisamy}

\begin{abstract}
Password-based authentication is one of the most commonly used methods for verifying user identities, and its widespread usage continues in virtual reality (VR) applications. As a result, various forms of attacks on password-based authentication in traditional environments such as keystroke inference and shoulder surfing, are still effective in VR applications. While keystroke inference attacks on virtual keyboards have been studied extensively, few efforts have developed an effective and cost-efficient defense strategy to mitigate keystroke inferences in VR. To address this gap, this paper presents a novel QWERTY keyboard called \textit{VRSafe} that is resilient to keystroke inference attacks. The proposed keyboard carefully introduces false positive keystrokes into the information collected by attackers during the typing process, making the inference of the original password difficult. \textit{VRSafe} also incorporates a novel malicious login detector that can effectively identify unauthorized login attempts using credentials inferred from keystroke inference attacks with high detection rate and minimal time and memory cost. The proposed design is evaluated through both simulation experiments and a real-world user study, and the results show that \textit{VRSafe} can significantly reduce the accuracy of keystroke inference attacks while incurring a modest overhead from a usability standpoint.
\end{abstract}

\ccsdesc[500]{Human-centered computing~Virtual reality}
\ccsdesc[500]{Human-centered computing~User studies}
\ccsdesc[500]{Human-centered computing~Keyboards}
\ccsdesc[500]{Human-centered computing~Text input}
\ccsdesc[500]{Security and privacy~Authentication}

\keywords{VR Security; Password-based Authentication; Keystroke Inference Attack Defense}


\fancyhead{}
\maketitle

\section{Introduction}

Virtual reality (VR) devices and applications are becoming increasingly prevalent. Studies show that over 171 million people globally use VR~\cite{VRStatReport2025}, and nearly 88\% of VR users use their devices multiple times in a month~\cite{Beyond_reality}. VR applications provide immersive experiences not only in gaming and entertainment, but also in many other domains including education, fitness, healthcare and engineering~\cite{Beyond_reality}. The widespread use of VR leads to increased exchange of sensitive information (e.g. passwords), making them attractive targets for adversaries.

Password-based authentication stands out as the most commonly used authentication method to date, however, extensive prior work has shown its vulnerability to keystroke inference attacks and shoulder surfing across various settings~\cite{yang2022wireless,sun2016visible,cronin2021charger,wang2024muki,sabra2020zoom,gupta2018side}. With the growing adoption of VR devices, these risks become more severe. VR applications provide a rich immersive experience to the user, enabling more stealthy attacks as users are generally less aware of their physical surroundings when using the VR devices. Many inference attacks only require placing an inconspicuous sensor (e.g. camera) and achieve high inference accuracy~\cite{yang2023towards,gopal2023hidden,al2021vr,luo2024eavesdropping,lee2023vrkeylogger,zhang2023s,slocum2023going,wu2023privacy}. These threat models exploit various information such as user's hand movement~\cite{gopal2023hidden}, gyroscope data~\cite{slocum2023going}, wifi-signal data~\cite{al2021vr} or acoustic signals \cite{luo2024eavesdropping}. Such attacks are highly
practical in the real world as they incur low cost and require minimal use of specialized knowledge. Prior efforts have also explored biometric-based approaches as an alternative to password-based authentication~\cite{soni2021keynet,luo2020oculock,shen2018gaitlock,boutros2020iris} such as using eye-tracking cameras to capture iris information in VR headsets~\cite{luo2020oculock,boutros2020iris}. Although biometric approaches have better resistance against keystroke inference attacks, they have not gained wide acceptance among the users. This is in part due to the lack of mature regulations and laws on how to store highly unique biometric data, as well as due to OS and hardware compatibility issues~\cite{stephenson2022sok} across different commercial VR products. 

In existing threat models, the risk of full password inference is underestimated. Though existing keystroke inference attacks can achieve high character-level guessing accuracy (often exceeding 95\%)~\cite{yang2023towards}, inferring a full password string is significantly more challenging due to the absence of inter-word context and limited intra-word semantics compared to inferring words in natural language texts (e.g. email content). While simple typographic correction tools~\cite{wu2023privacy} may yield reasonable improvements for general text inference, such tools offer little benefit for reconstructing passwords. Designing or evaluating defense mechanisms based on such a threat model can lead to overly optimistic conclusions, as these methods fail to capture the strategies that a practical adversary would adopt. To build a more robust and practical defense, in this work, we adopt a stronger threat model in which the attacker integrates advanced password guessing techniques~\cite{wang2023pass2edit,yang2025targeted,pal2019beyond,nosenko2023password} into the inference process, rather than relying on simple correction tools.

Although there have been many research efforts on studying keystroke inference attacks in VR environments, countermeasures on this topic have not received adequate attention yet. Some studies briefly discuss ideas for potential defenses~\cite{yang2023towards}, others have proposed and evaluated defense solutions based on obfuscating user inputs using randomized keyboard layouts~\cite{maiti2017randompad} or by repositioning the UI components~\cite{wan2024analysis}. However, such techniques increase the usage difficulty and impose additional physical and cognitive loads to the user.

There are several challenges involved in building an effective countermeasure against keystroke inference attacks. First, many attack models can reliably detect the precise moments when keys are pressed (e.g.~\cite{gopal2023hidden,yang2023towards,slocum2023going,lee2023vrkeylogger}). A strong defense method should therefore remain effective even under such worst cases, where the adversary can perfectly identify every key press event made by the user. Second, when an attacker attempts to log in with an incorrect but closely matching password, it is inherently difficult to detect the source of the password leakage. A comprehensive countermeasure may detect such malicious login attempts in addition to mitigating them. Finally, maintaining a high level of usability in the defense solution design remains a critical challenge.

In this paper, we propose \textit{VRSafe}, a security solution designed to enhance the security of password inputs entered through a standard QWERTY keyboard in virtual reality. The proposed approach carefully injects fabricated noise into the typing process, making the inference of the original password harder for the attacker. \textit{VRSafe} also incorporates a novel detection mechanism to identify whether an incorrect login attempt is a direct result of a keystroke inference attack. To the best of our knowledge, \textit{VRSafe} is the first approach to alert the user when an attacker attempts to login using their inferred keystroke information. Our evaluation using public leaked password datasets and an IRB-approved user study demonstrates that \textit{VRSafe} is a secure, practical and user-friendly approach for individuals with higher security and privacy needs in VR environments.

\section{Background and Related Work}
\label{sec:related work}
In this section, we discuss the existing works on keystroke inference attack and defense in VR.

Keystroke inference attack, as the name suggests, tries to reconstruct the user's input sequence on the keyboard through side-channel information emanated from typing (e.g. sound, motion). Over the past decade, keystroke inference attack has been widely investigated, prompting the use of many different techniques to leverage these side-channel signals. Recent work~\cite{gopal2023hidden,ling2019know,yang2023towards} have demonstrated the feasibility of reconstructing sensitive texts through analyzing hand movements captured by videos, highlighting the critical need for enhanced privacy and security protection. It is also possible to use the hand tracker camera of an Augmented Reality Head-Mounted Display (AR-HMD)~\cite{meteriz2022keylogging}  to recover passwords similar to the video-based attack in the VR environment. Wang et al. ~\cite{wang2024gazeploit} find that gaze motions observed in videos can also leak typing sequence information. In addition to video-based attacks, attacks that exploit other channels such as audio, radio frequency (RF) or infrared signals generated during typing also pose serious threats. \textit{VRecKey} ~\cite{ni2024non} discloses an attack in which Infrared Remote (IR) signals emitted from VR controllers can be represented in heat map to reveal controller positions during typing in order to infer keystroke information. The attack presented in \cite{al2021vr} uses Channel State Information (CSI) of Wifi signals from routers placed near the victim. The information exhibits unique patterns as users move the controller to type different keys, helping the attacker to identify the keystrokes. Another class of attacks uses the acoustic signals emitted by the VR devices. \textit{Heimdall}~\cite{luo2024eavesdropping} captures subtle variance in acoustic information created by the controller in different key positions using their customized directional microphone. Prior works have also shown that attackers can utilize Inertial Measurement Unit (IMU) data such as combination of position, orientation and velocity~\cite{slocum2023going,luo2022holologger,wu2023privacy} or memory usage in the victim's device~\cite{zhang2023s} to perform keystroke inference.

In contrast to the widespread prior efforts in developing various attacks, models to mitigate or to prevent inference attacks in VR environments are much less explored. In practice, very few methods can balance ease of use with strong performance in defending against the inference attacks. The defense solution outlined in \cite{yang2023towards} suggests placing physical screens to block the external line of sight. The approach presented in \cite{10.1007/978-981-99-8024-6_13} replaces the normal 2D QWERTY keyboard with a curved, arc-shaped layout in AR to introduce variation in key spacing. Another approach to safeguard against keystroke inference attacks is to introduce randomness. For instance, the solution presented in \cite{maiti2017preventing} proposes projecting a random layout keyboard, and \cite{li2020designing} generates a random mapping between alphabets and keypad keys to secure password entry in smart wearable glasses. The authors in \cite{wan2024analysis} propose repositioning the UI components (e.g. cursor, keyboard) after each click so that an adversary cannot reliably map observed movements to specific keys.

Overall, current solutions often adopt radical changes to the conventional keyboard layouts or input designs, and some rely on extra external protection devices. Such approaches introduce substantial learning overhead and increase the difficulty of use for end users. In contrast, the research presented in this paper aims to fill this gap by developing a usable and secure VR keyboard architecture that can withstand video-based keystroke inference attacks and yet provide a highly usable VR experience to the user.

\section{\textit{VRSafe} Design}\label{Overview}

In this section, we present \textit{VRSafe}, a secure keyboard in virtual reality that provides word-level protection for user password input against keystroke inference attacks. We first introduce the threat model used in the design of \textit{VRSafe} and then describe the strategies employed for achieving better security and balancing the overhead incurred while using the keyboard. Finally we present our proposed mechanism for detecting malicious login attempts using credentials stolen through keystroke inference attacks.

\subsection{Threat Model}

\subsubsection{Keystroke Inference}
Keystroke inference threat models can employ a range of sensors, each with its own tolerance to noise. Video-based attacks often use ordinary digital cameras with no special configuration~\cite{yang2023towards,ling2019know,luo2022holologger,gopal2023hidden}, and videos are generally more robust to noise. To ensure reproducibility and fair performance comparisons with prior work, we adopt a video-based attack in our study. Specifically, we follow the attack model proposed in~\cite{gopal2023hidden}, where the adversary captures the VR user's hand movements with a camera placed in the surrounding environment. We believe such a threat model is quite practical and is within the capabilities of most real world adversaries. In this model~\cite{gopal2023hidden}, the victim is assumed to interact with the keyboard using hand tracking features provided by VR headsets for typing. The captured video frames are processed using the hand landmark detection framework in MediaPipe~\cite{zhang2020mediapipe} to extract coordinates of hand joints. By leveraging unique joint patterns exhibited during typing, such as relative joint positions and velocities, the model infers both the spatial position of the virtual keyboard plane and the final keystroke predictions. To better model the attacker's ability, we make three more assumptions, each with a brief rationale: (i) The attacker can correctly identify every keystroke in the corresponding video clips. While raw video data may include false positives and false negatives, techniques like denoising and clustering can significantly reduce such detection errors, leading to a high true positive detection rate. (ii) The attacker can access any publicly available information within the application, but does not have access to private data of the legitimate user. Public data like keyboard layouts and language preference are easily obtainable, whereas private data such as screen display content are protected and inaccessible without privileged permissions. (iii) We also assume that the attacker is allowed at most \(k\) guessing attempts. This is consistent with most web applications, which allow only a small number of password guesses to be made before the system locks the user for suspicious action.

\subsubsection{Password Guessing Refinement}

Although existing keystroke inference attacks can achieve high character-level accuracy, end-to-end word guessing accuracy still has significant room for improvement. Prior work has used simple typo correction tools such as grammar and spelling correction in Google Doc~\cite{wu2023privacy} on texts reconstructed from keystrokes, which underestimates the capabilities of real world attackers. In contrast, we find that using targeted password guessing models can improve the full password guessing accuracy, providing a more realistic assessment of defense effectiveness. Particularly, Pal et al.~\cite{pal2019beyond} proposed a sequence to sequence (seq2seq) targeted password guessing model using recurrent neural network (RNN) which is trained on users’ leaked passwords from a breached site to guess their passwords on another uncompromised site. Wang et al.~\cite{wang2023pass2edit} introduced \textit{Pass2Edit}, which further improves this idea and currently achieves state-of-the-art performance. By inputting passwords inferred from keystrokes into such models, one can generate a list of plausible passwords, and one of which could correspond to the actual password.

{\subsection{Design Overview}

\textit{VRSafe} consists of two modules, Keystroke Noise Injection Keyboard and Inference Attack Detector, as shown in Figure~\ref{fig:systemflow}. Users interact with \textit{VRSafe} keyboard in a way similar to normal virtual keyboards, with only a small number of additional actions that intentionally introduce noise into the user's password input. \textit{VRSafe} can differentiate the noise from the actual password, maintaining the original password unchanged while storing a "noisy" version of the password. Upon clicking the login button, both the original and "noisy" password will be sent to the server. The server verifies the original password against the stored credentials in the database and only if the verification succeeds, it adds the corresponding “noisy” password to the detector. The shaded region in Figure~\ref{fig:systemflow} represents a keystroke inference attacker who covertly places a camera to record the user’s interactions with the VR headset, including the additional actions required by \textit{VRSafe}. Based on these recordings, the attacker attempts to infer keystrokes and launch a malicious login attempt. However, the attacker can only recover the “noisy” password from the recorded video, and any such login attempts using the “noisy” passwords will be detected by the server.

\begin{figure}[htbp]
    \centering
    \includegraphics[clip=true, trim={0 0 00 00}, width=\linewidth, scale = 1]{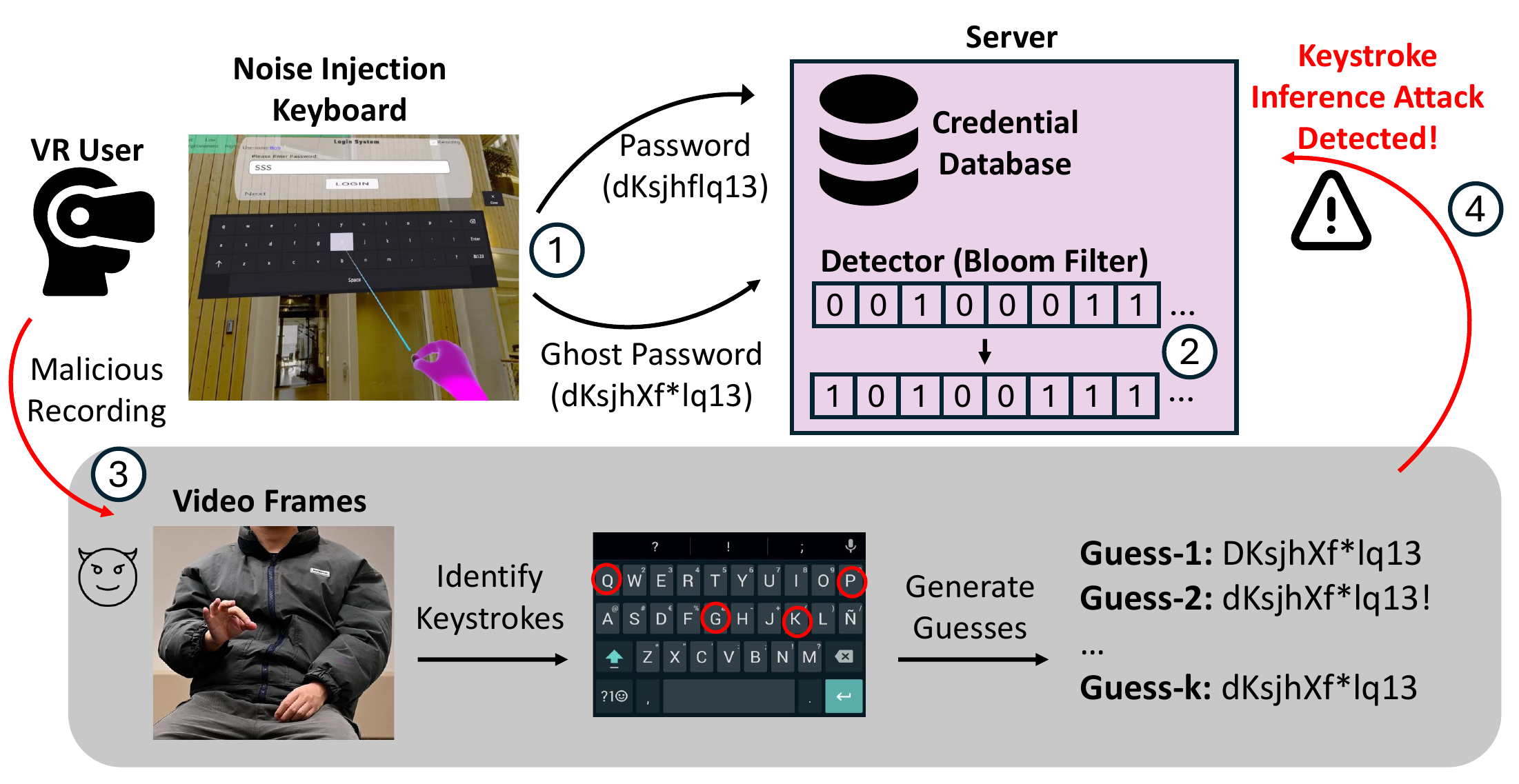}
    \caption{\textit{VRSafe} architecture. (1) Normal user interacts with \textit{VRSafe} keyboard, the original password should remain unchanged for successful authentication and "noisy" password will be forwarded to detector. (2) Server adds the "noisy" password to detector if the original password matches the record in database(i.e. a legitimate login). (3) A keystroke inference attacker captures the user's keystroke on \textit{VRSafe} keyboard and attempts to login. (4) The detector raises an alarm if credential exists in the detector.}
    \Description{Overview of our proposed \textit{VRSafe} solution.}
    \label{fig:systemflow}
\end{figure}

\subsection{Keystroke Noise Injection in VR Keyboard}\label{sec: Keystroke Noise Injection Keyboard}

We first explain the design rationale behind our approach of injecting fabricated noise to mitigate keystroke inference attacks and present the technical details of our design. The keystroke noise injection in \textit{VRSafe} is based on insights from recent work demonstrating that even for attack models with high accuracy~\cite{yang2023towards}, certain behaviors significantly affect the accuracy: a participant pressing down towards a key, hesitating and then retracting the finger(s) before hitting, can lead to large amounts of false positive keystrokes. Drawing insights from this observation, we explore this phenomenon and propose our keystroke noise injection mechanism. Instead of accidentally "false pressing a key" or asking users to type something wrong occasionally~\cite{yang2022wireless}, \textit{VRSafe} leverages an adaptive mechanism that prompts the user to insert fabricated characters (noise) based on the current input string and previously injected noise. An attacker observing the hand movements of the VR user will not be able to differentiate the true intended inputs from fabricated ones, resulting in incorrect password inference. From a password protection standpoint, this process can be viewed as generating an augmented password \(S'\) derived from the original password \(S\). There is a constraint in this process: \(S\) must be a \textbf{subsequence} of \(S'\), otherwise we will lose content in the original password. At the same time, the key press signals triggered by fabricated characters are processed differently and will not be forwarded to the keyboard output function (Fig.~\ref{fig:VRSafeDesign}). This ensures the integrity of the normal login process. We refer to the characters that exist in user's original/real password as \textbf{real characters}, the additional fabricated characters prompted by \textit{VRSafe} as \textbf{ghost characters} and the augmented password as \textbf{ghost password}.

\begin{figure}[htbp]
    \centering
    \includegraphics[clip=true, trim={0 0 0 0}, width=0.8\linewidth]{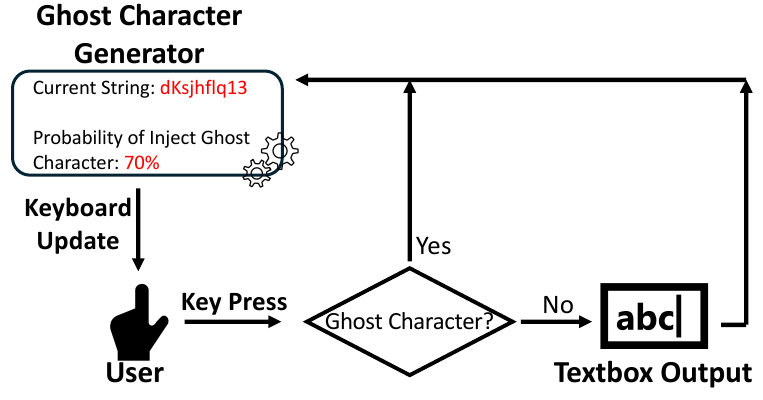}
    \caption{\textit{VRSafe} will maintain a state variable to indicate whether the input is a ghost character or a real character, and only real characters will be forwarded to the textbox.}
    \Description{Typing process of a user using \textit{VRSafe} keyboard.}
    \label{fig:VRSafeDesign}
\end{figure}

\noindent\textbf{Ghost Password Generation.} Many models such as RNN-based seq2seq model~\cite{sutskever2014sequence}, transformer based LLM models~\cite{vaswani2017attention,achiam2023gpt} are capable of generating a similar password sequence based on a given password. However, they are inadequate for this task as they cannot guarantee the subsequence constraint (i.e. original password \(S\) is a subsequence of ghost password \(S'\)) is satisfied on every input/output pair. To address this limitation, we need to employ a method that can reliably embed the characters of the original password into the ghost password. In \textit{VRSafe}, we adopt a pointer-based method with two actions: (i) \textit{copy} action, which appends the current pointed character in the original password to ghost password, and (ii) \textit{inject} action, which appends a random ghost character. By progressing the pointer from the first to the last character of the original password, this approach ensures that the subsequence constraint is always satisfied. To decide between the two actions, we model the process as a finite sequence of Bernoulli trials: with probability \(P\), the model performs an inject action and with a probability \(1-P\), it performs a copy action. Furthermore, to prevent the ghost password from becoming excessively long or identical to the original password, we impose a maximum limit on the number of consecutive ghost characters and a minimum requirement on the total number of injected ghost characters.

Like words in natural languages, many passwords also contain semantics, especially in human-chosen passwords. If we inject a ghost character into an improper position, the adversary can easily spot the ghost character and remove it. Conversely, if a password already has a high entropy, adding many ghost characters provides little additional benefit. Therefore, the ghost password generation process should dynamically choose which action to take based on the current context. To address this, we train a small RNN-based model as a meter to evaluate the "randomness" of the current string, and adjust the injection probability \(P\) accordingly. The model consists of a single-layer Gated Recurrent Unit (GRU) with a hidden size of 64, and input tokens are represented with embeddings dimension of 16. The GRU output is then fed into a fully connected (FC) layer and the final output is passed through a sigmoid activation to return a value between 0 and 1. For training, we use a publicly leaked dataset, namely Compilation of Many Breaches (COMB) dataset~\cite{COMB_news}. We label passwords from publicly leaked password dataset with 0 and machine generated passwords with 1. We note that not all passwords in public leaked dataset are human-chosen, but there is no consensus on how to reliably distinguish them, since passwords do not have "correct spelling" or linguistic rules like natural language words. Nevertheless, leaked password datasets remain the most representative and widely used source of human created passwords, making them a reasonable choice for approximating the human-chosen class in this case. To exclude extreme outliers, we remove passwords that exceed 30 characters or only contain hexadecimal characters, since they are more likely to be hashes than passwords. The ghost password generation algorithm (Algorithm~\ref{alg:update_p}) starts with an initial injection probability \(p_0\). After each action, the current ghost password will be forwarded to the meter model, the return value will be smoothed through exponential moving average (EMA) and compared with randomness level \(r\). If it is smaller {than \(r\), meaning that the current string is not adequately random, then \(p\) will increase \(\Delta p\), otherwise \(p\) becomes smaller.

\begin{algorithm}[h]
\caption{Adaptive Ghost Character Injection}
\label{alg:update_p}
\begin{algorithmic}[1]
\Require smoothing factor $\alpha$, randomness level $r$, injection probability $p_0$, step size $\Delta p$, password $pwd$
\State Initialize $p \gets p_0$, $r_{\text{EMA}} \gets r$, ghost\_pwd $\gets ""$, $pos = 0$
\While{ $pos < \text{len}(pwd)$}
    \If{random() $< p$}
        \State \textit{Inject}(ghost\_pwd)
    \Else
        \State \textit{Copy}(ghost\_pwd, pwd, pos)
        \State $pos \gets pos + 1$ \Comment{moves forward to next character}
    \EndIf
    \State $\hat{r} \gets \textit{Eval}(\text{ghost\_pwd})$ \Comment{Evaluate randomness of current string with meter model.}
    \State $r_{\text{EMA}} \gets (1 - \alpha) \cdot r_{\text{EMA}} + \alpha \cdot \hat{r}$
    \If{$r_{\text{EMA}} < r$}
        \State $p \gets p + \Delta p$
    \Else
        \State $p \gets p - \Delta p$
    \EndIf
\EndWhile
\State \Return ghost\_pwd
\end{algorithmic}
\end{algorithm}

\noindent\textbf{Choosing Ghost Characters.} We further elaborate the noise injection process (\emph{Inject()} in Algorithm~\ref{alg:update_p}) in the generation of ghost passwords. A straightforward baseline approach is to uniformly choose a ghost character from an alphabet set of all legal characters. However, a selection based on uniform distribution remains vulnerable to simple human inspection. For instance, suppose that the original password contains a date of birth, and if an alphabetic letter is added as a ghost character in between, then the attacker can easily exclude this letter from the inferred text. In order to make the ghost character look similar to a real character, we select ghost characters using simple language models that can capture linguistic connections between characters to make the process context-aware. Unlike neural networks that often require significant computational resources and introduce additional latency, the Markov model offers a lightweight and efficient alternative for ghost character selection. By definition, a \(k\)-order Markov model assigns a probability distribution over the next character conditioned only on the preceding \(k\) characters:

\[
\begin{split}
P(x_i \mid x_{i-1}, x_{i-2}, \dots, x_1) 
&= P(x_i \mid x_{i-1}, x_{i-2}, \dots, x_{i-k}) \\
&= \frac{\text{count}(x_{i-k}, x_{i-k+1}, \dots, x_i)}{\sum_{c \in \Sigma} \text{count}(x_{i-k}, x_{i-k+1}, \dots, x_{i-1}, c)}
\end{split}
\]

\noindent where \(count(x_{i-1},x_i)\) refers to the number of occurrences of string \(x_{i-1}x_i\) in a given dataset. We test the Markov model using the configuration used in ~\cite{ma2014study} with different orders and find that the 3-order Markov does not incur any noticeable delay in the ghost character generation process and yet provides sufficient context-awareness and randomness.

\noindent\textbf{Implementation.} The keyboard in \textit{VRSafe} is built using Unity based on a publicly available QWERTY keyboard template from the MRTK toolkit by Microsoft\cite{MRTK}. While entering the password, if the next character is a ghost character, the keyboard will disable all other keys on the keyboard except the ghost character and notify the user via text prompts, ensuring the user's next action injects the intended noise. After typing the ghost character, the keyboard layout will be restored to normal QWERTY and the user can continue typing. In a normal keyboard, every key fired by the user click will be appended to the textbox. We add a private variable to track whether each character is a real or a ghost character, and only real characters are forwarded to the text output. The ghost password is stored alongside the real password in the textbox GameObject, ensuring that it persists even if the keyboard is closed. We also create a login interface together with the \textit{VRSafe} keyboard as a VR application using Unity 2022.3.24f1 and deploy it on Meta Quest2 for the user study. The screenshot of the application is shown in Figure~\ref{fig:VRSafeInference}.

\begin{figure}[htbp]
    \centering
    \includegraphics[clip=true, trim={0 0 0 0}, width=0.75\linewidth]{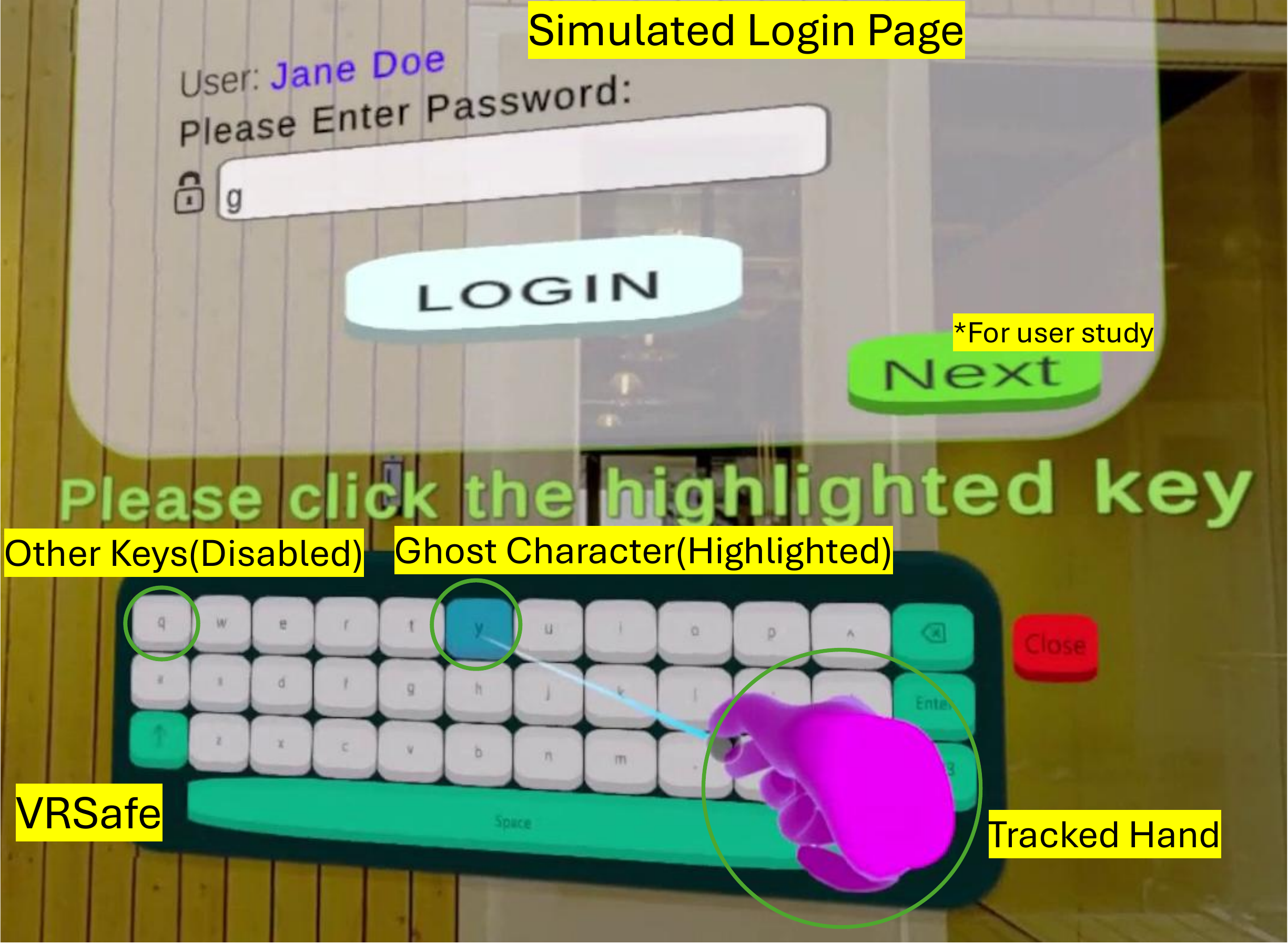}
    \caption{A screenshot when the user needs to type a ghost character. All other keys except the ghost character will be disabled, and the user will be notified via text prompts.}
    \Description{A screenshot of our \textit{VRSafe} keyboard application.}
    \label{fig:VRSafeInference}
\end{figure}

\noindent\textbf{Overhead.} Ghost characters enhance the security of the original password, however, they also incur longer text entry time. Therefore, it is critical to carefully consider the overhead introduced in the process. Although entry time can be influenced by various factors such as the user’s experience with the device and typing proficiency, we focus on reducing the overhead introduced by the additional characters, and we further discuss the subjective factors of users' perceived overhead in the user study (Section \ref{userstudy}).}

\begin{figure}[htbp]
    \centering
    \includegraphics[clip=true, trim={00 0 00 00}, width=0.65\linewidth]{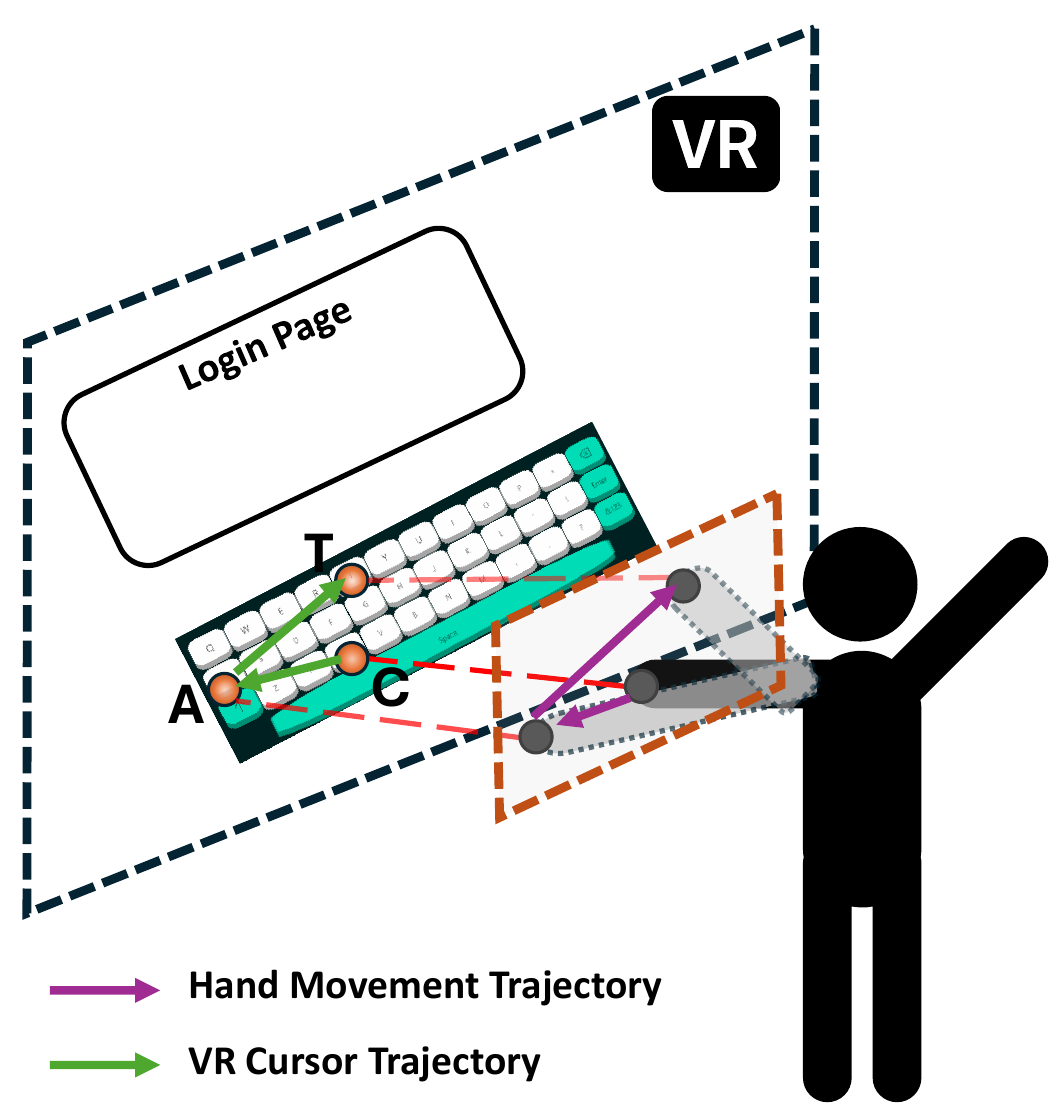}
    \caption{The green arrows depict the cursor’s trajectory in keyboard space, and the purple line denotes the user’s hand trajectory in the physical world for typing word "CAT" using virtual keyboard.}
    \Description{Cursor and hand movement trajectory of typing word "CAT".}
    \label{fig:distance}
\end{figure}

In VR applications, the cursor is often bound to handheld controllers, and interactions with components require moving the cursor to the target position. Typing on a virtual keyboard typically involves two steps: moving the cursor to the target key and pressing the button on the controller to select, and then this process repeats for each character. We can see that a substantial portion of entry time is spent on the controller movement. Meanwhile, to provide a realistic user experience, the ratio between cursor movement in virtual space and physical controller movement in the real world is often close to one~\cite{Meta_locomotion}. For example, the hand and cursor movement trajectories for typing the word “CAT” on a virtual keyboard are illustrated in Fig.~\ref{fig:distance}. Thus, the entry time of a given password can be approximated by the total distance traversed by the controller, or equivalently by the cursor. Suppose \(C_i = (x_i,y_i)\) denotes the coordinates of the $i^{th}$ character in the password on the keyboard plane, then the total distance traversed by the cursor for the entering the password is computed using the following formula, where \(\|\cdot\|_2\) denotes Euclidean distance:

\[
d_{sum} = {\sum_{i=2}^{n} \| {C}_i - {C}_{i-1} \|_2}
\]

\noindent Based on the above formula, the overhead can be calculated as the difference in the distance between the original password and the ghost password. Our objective is to reduce the overhead without significantly compromising the enhanced security achieved by \textit{VRSafe}. We can either reduce the number ghost characters or use weighted selection instead of uniform random choice. The former can be adjusted by choosing different randomness levels discussed in Algorithm 1. Here, we primarily focus on the latter. In natural languages, certain bi-grams occur with similar frequencies, yet their corresponding cursor traversal distances on the QWERTY keyboard can differ substantially. For instance, the bi-gram "PA" shows up 10963 times and "PO" appears 11965 times in one million English words~\cite{solso1979bigram}. This indicates that both bi-grams are equally plausible from a linguistic standpoint, selecting either “A” or “O” as a ghost character offers no meaningful advantage to an attacker. However, typing "PO" results in a much smaller distance cost than "PA". Therefore, we can choose ghost characters that are located closer on the keyboard to reduce the moving overhead and yet achieve a similar protection performance. We implement this constraint in two ways, namely hard constraint and soft constraint. With hard constraint, we remove the characters that are unacceptably far from the preceding character by defining a threshold distance \(\tau\) and assigning zero probability to any candidate that exceeds the threshold. In contrast, the soft constraint does not explicitly eliminate distant characters from the candidate set. Instead, it adjusts their probabilities of being selected based on the distance. We compute the distance \(d\) between each candidate ghost character \(c_i\) and preceding character \(c_{i-1}\), and adjust the selection probability using a softmax weighted function where closer characters receive higher probabilities:

\[
P'(c_i) \;=\; 
\frac{
P(c_i) \, \exp\!\left( -\lambda \, d(c_i, c_{i-1}) \right)
}{
\sum\limits_{j} P(c_j) \, \exp\!\left( -\lambda \, d(c_j, c_{i-1}) \right)
}
\]

\noindent Higher \(\lambda\) or lower \(\tau\) will both result in smaller overhead with a more compact key distribution. However, it might increase the risk of weakening the enhanced security. We discuss the performance of \textit{VRSafe} for different values of \(\lambda\) and \(\tau\) in Section \ref{Experiment}.

\subsection{Inference Attack Detector}

Although the proposed noise injector increases the difficulty of obtaining the original password, an attacker may still correctly guess the password given enough attempts. In this case, the total number of brute-force attempts required to infer the correct original password from the ghost password is approximately \(k \times 2 ^ n\), where \(k\) is the number of inferred ghost password candidates and \(n\) is the length of the ghost password owing to the subsequence relationship between the ghost password and the real password. Thus, it is crucial to add a detection mechanism to identify whether an account is being actively targeted or guessed. By identifying targeted guessing attempts early, the service provider can notify the user to limit potential damage. We add a checker at the server side inspired by the honey checker proposed in~\cite{DBLP:conf/ccs/JuelsR13}. The approach behind honey checker is to create false passwords (also called "honeywords" or "decoy passwords") that are similar to the user's real password. All password hashes are stored in the credential database, and the server knows which one is authentic. An adversary who inverts the hashes will obtain multiple password candidates of each user, and if the attacker attempts to authenticate with a decoy password, the server can immediately detect and flag the unauthorized attempt. Typically, this approach is employed to defend against attacks where password hash files from credential databases are compromised by data breaches. In our context, the attacker has knowledge of the ghost password which they believe is the original password of the user, and this password can also be considered as a "decoy password". Therefore, for better illustration, we refer to the password used for detecting keystroke inference attacks as "honeyword". If the adversary attempts to login with the ghost password, we can infer that a keystroke inference attack has previously taken place. We draw this inference because neither the legitimate user nor other types of attackers would use the ghost password for authentication, and only a keystroke inference attacker possesses knowledge of the ghost password. 

To the best of our knowledge, no prior work has explored this setting, and our approach provides a novel detection mechanism specifically for keystroke inference attacks. Some research also explored alarm raising attackers~\cite{DBLP:conf/ndss/WangR24,DBLP:conf/uss/HuangBR24,DBLP:conf/sp/0002ZDSH22} where the goal is to trigger false alarm than gaining unauthorized access. We argue that keystroke inference attackers are unlikely to adopt this strategy for the following reason: if an attacker already acquires user password from keystroke inference attack, raising alarm would not give them any profit about user's sensitive information, and the password would expire soon since the compromised user will be notified by the server to change the password right away. As a result, a keystroke inference attacker who is interested in user's sensitive data would always implement a false negative attack (i.e. try to login without being noticed).

\noindent\textbf{Honeyword Selection and Storage.} Without any doubt, the ghost password should be included as a honeyword, as it is highly likely to be selected by an attacker in a login attempt. To broaden the detection surface, we should include as many honeywords as possible, however, introducing too many honeywords may greatly increase verification time and impair the performance. To maximize detection rate while using a reasonable number of honeywords, we rank all possible guesses in descending order of probability from the password guessing model and choose the top \(n-1\) guesses excluding the ghost password and the original password, making \(n\) honeywords in total. As for false alarms caused by the user due to typographic errors, since the ghost password has a requirement for the minimum number of injected characters, the likelihood that a user's typo accidentally matches the ghost password should be minimal.

We now discuss how to store the honeywords at the server. We note that the ghost password is generated from a probabilistic model which varies across login sessions, and the honeywords are derived based on it. As a result, each time a user logs in using \textit{VRSafe}, we might get a different honeyword set and the total number tends to increase as the user continues to log in multiple times over time. To reduce the storage overhead associated with the growing number of honeywords, we use bloom filter (BF)\cite{bloom1970space}, a compact data structure that tests whether an element is a member of a set with possible false positives but guarantees that there are no false negatives. Using BF has two advantages over storing honeywords as individual hashes. First, it provides insert and lookup operation with constant time $O(k)$ regardless of the number of elements already inserted, where \(k\) is the number of hash functions of BF defined during initialization. Second, the space cost of BF is fixed and it does not grow with the number of stored elements. However, BF also has its limitations. For a given BF consisting of \(k\) hash functions and size of the bit array \(m\), there would be a maximum expected elements \(n\) given a certain false positive rate (FPR) \(p_{fp}\):
\[
p_{fp} \approx \left(1 - e^{-\frac{kn}{m}}\right)^k
\]
\noindent At the same time, the bloom filter does not support deletion operation, which means that once the number of inserted elements approaches the expected maximum \(n\), we can only rebuild a new filter, otherwise the FPR will become higher than expected. Therefore, we need to have a good estimate on the expected growth trend of the password logins by the user in order to avoid rebuilding the BF frequently. According to a study on web password habits~\cite{florencio2007large}, on average, a user logs into a website approximately 3.22 times per day (i.e. roughly 1,000 times per year). Such estimates can be used to determine the expected interval (e.g., 6 or 12 months) for rebuilding a BF and the expected maximum number of elements.

\section{Experiments}\label{Experiment}
We evaluate \textit{VRSafe} through both simulations and a real-world user study approved by the Institutional Review Board (IRB). Before presenting the experiment results, we first describe our experiment setup.

\subsection{Experiment Setup}\label{setup}

\begin{table}[H]
\small
\caption{Summary of training time and iterations of models used in this paper.}
\begin{tabular}{lcc}
\hline
Model          & Epoch            & Training Time (Total)  \\ \hline
Pass2Edit      & 3                & 24hr                 \\ 
Randomness Meter& 10               & 10hr                 \\ 
3-Order Markov & - & 2hr       \\ 
\hline
\label{tab:training time}
\vspace{-10pt}
\end{tabular}
\end{table}

\begin{table*}[]
\caption{Data cleaning result of leaked password dataset.}
\label{tab:Dataset}
\centering
\begin{tabular}{ccccccc}
\hline
\bf{Dataset} &  \bf{Leaked Time} & \bf{Raw}&  \bf{Non-ASCII} &  \bf{Empty} &  \bf{Removed(\%)} &  \bf{Cleaned} \\ 
\hline
COMB        & Feb. 2021 &  3,279,064,312 &  14,827,020 &   187,089   &  4.6   &  3,264,050,203
  \\ \hline
\end{tabular}
\end{table*}

We evaluate \textit{VRSafe} based on real leaked password datasets namely COMB datasets. The Compilation of Many Breaches (COMB) dataset was firstly leaked in Feb. 2021 on a popular online hacking forum~\cite{COMB_news}. The dataset includes the largest volume of recently leaked passwords from various websites.~\footnote{The dataset was collected from publicly accessible websites for research purposes only. We intentionally omit the source URL of the leaked dataset to limit further exposure.} Consistent with prior work\cite{wang2023pass2edit,pal2019beyond}, we remove entries that contain non-ASCII printable characters and empty rows. The cleaned dataset is described in Table~\ref{tab:Dataset}. All models are trained on Google Colab platform using T4 GPU and High-RAM mode. For both training and testing, we choose passwords between 5 and 30 characters in length within the dataset, as passwords outside this range are either too weak or unlikely to be human chosen passwords. The targeted password guessing model follows the setup in~\cite{wang2023pass2edit} except the data size. With the configuration data size used in ~\cite{wang2023pass2edit}, it results in more than 7 days of training time per epoch for large datasets. To ensure practical training efficiency, we instead sample 50 million password pairs. For the randomness meter model in Algorithm~\ref{alg:update_p}, we train on 5 million real passwords and an equal number of randomly generated passwords with the same length distribution for 10 epochs. The Markov model for ghost character selection is trained on 5 million passwords. Training time for all the models is shown in Table~\ref{tab:training time}.

\subsection{Experiment Results}\label{Performance}
We conducted a series of experiments to evaluate the factors that affect the guessing accuracy and overhead. We also evaluate accuracy and resource cost of our detection mechanism.

\noindent\textbf{\#1: Targeted Guessing vs. Simple Typo Correction.} We first illustrate that evaluating the effectiveness of our defense under the password guessing model is a more realistic setup. To simulate passwords inferred from keystrokes, we randomly sample 20,000 passwords in the test set. Specifically, each character is replaced with one of its adjacent keys on the QWERTY layout with a probability of 5\%, which aligns with a character level inference accuracy of approximately 95\% in state-of-the-art studies~\cite{yang2023towards}. We then copy all passwords into a Google Doc, refining passwords with "spelling and grammar" tools. Some suggestions will split the word into phrases (e.g. "iloveyou" to "i love you"), we only accept the suggestions that retain a single word. At the same time, we input the passwords into the guessing model and pick the top-1 guess of each password. We find that the refined passwords from Google Doc have similar performance (64.55\%) compared to targeted password guessing model (64.69\%) when no ghost characters are injected, but it quickly falls behind once a small amount of noise is introduced (27.52\% vs. 32.78\% with an injection probability \(P = 0.1\)). In real-world cases, it is also likely to have false negative or false positive keystrokes, which will result in larger divergence between inferred passwords and original passwords. Therefore, using the password guessing model can more accurately reflect the performance of our defense against realistic attackers.

\noindent\textbf{\#2: Accuracy Analysis.} We sample passwords in different categories to simulate users with different password choosing habits. First, each character in a password can be classified as either digits, letters or symbols~\cite{weir2009password}, and prior work has shown that the number of classes is one of the key factors affecting password guessing performance~\cite{melicher2016fast,wang2023pass2edit,kelley2012guess}. We categorize passwords based on the number of classes they consist of, denoted as \text{Class-X} where $X$ denotes the number of classes of characters in the password. For instance, the password string "Jamesbond007" belongs to \text{Class-2} passwords. In addition, password length is another important factor that influences password guessing performance, with longer passwords generally being more difficult to guess accurately. In ~\cite{wang2023pass2edit}, the researchers found that their model outperformed prior approaches for passwords with lengths between 10 and 16 characters. Based on this, we categorize passwords into Short (\(< 10\)), Medium (\(10-16\)) and Long (\(> 16\)) passwords to provide comprehensive evaluations of our defense method. We input passwords to our noise injection algorithm using \(p_0 = 0.5, \Delta p = 0.05, \alpha = 0.1\) at various randomness level \(r\). Then we forward the ghost passwords to \textit{Pass2Edit} password guessing model to determine the final adversary's guesses. Fig.~\ref{fig:password2x3} shows password guessing accuracy under different password lengths and number of classes at different number of guesses (\(10,100,1000\)), which are common numbers evaluated in prior research~\cite{wang2023pass2edit,pal2019beyond,wang2016targeted}. The x-axis represents different randomness levels \(r\) as described in Algorithm~\ref{alg:update_p} and y-axis represents guessing accuracy under top-k guesses. Overall, using the Markov model in selecting ghost characters results in a better performance compared to uniform selection, especially for long passwords (see Table~\ref{tab:augmentation_accuracy}). The password accuracy of Class-2, Long passwords using Markov model achieve 5\% to 15\% lower accuracy compared to uniform selection under the same setting when 1,000 guesses are allowed across all randomness levels (Fig.~\ref{fig:password2x3}(e)). With different number of allowed guesses, we observe that the adversary’s guessing accuracy has an approximately linear relationship with the logarithm of the number of guesses, and the accuracy under 10 guesses is only around 20\% (see Fig.~\ref{fig:log_guess_acc}). When the number of guesses increases, an attacker typically needs to generate ten times as many candidates in order to achieve approximately twice the accuracy, and the improvement is even smaller for more complex passwords.

\begin{figure}[h]
  \centering
  \includegraphics[width=0.38\textwidth]{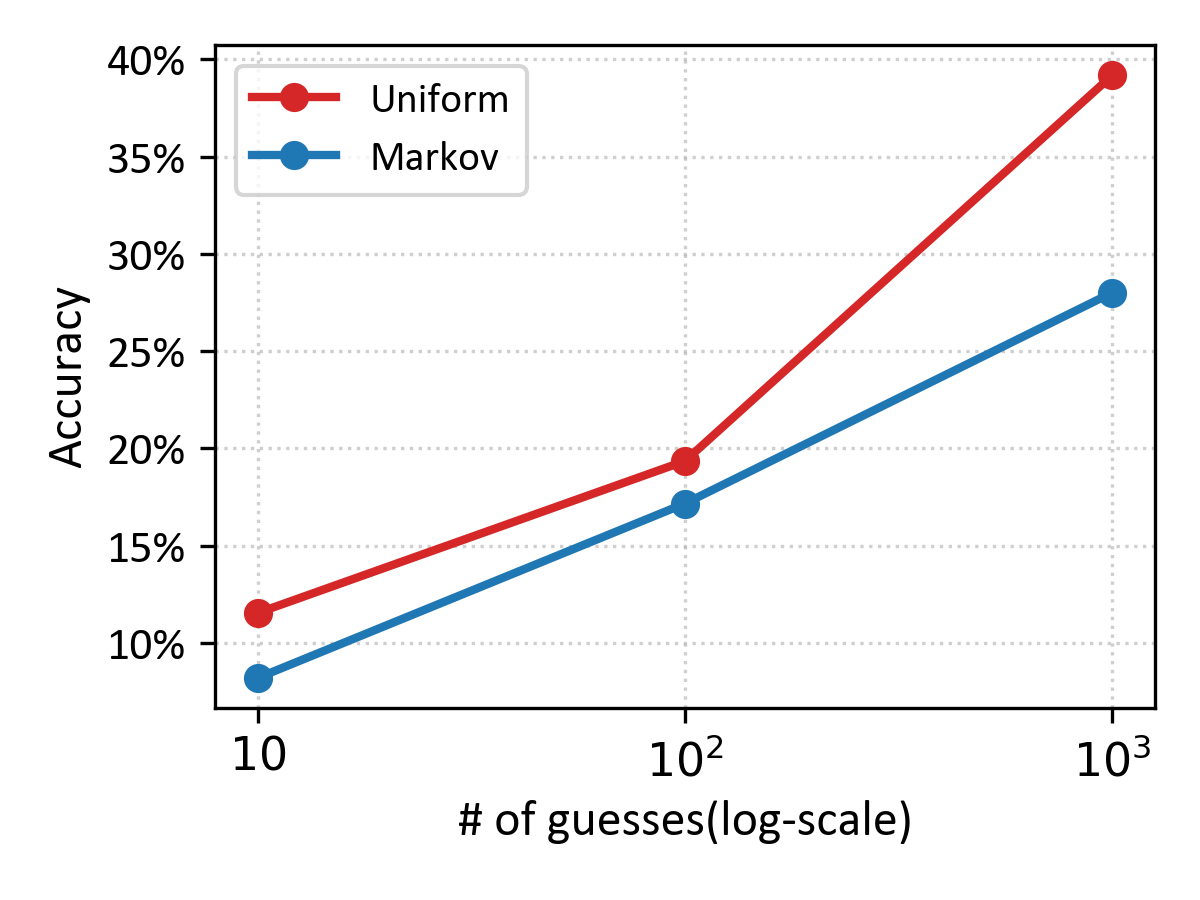}

  \caption{The average guessing accuracy for randomness level in [0.4,0.7] for both Markov and Uniform ghost character generation under various \# of guesses. Take Fig~\ref{fig:password2x3}(e) as an example.}
  \Description{Guessing accuracy with log scale of \# of guesses.}
  \label{fig:log_guess_acc}
\end{figure}

\begin{figure*}[t]
  \centering
  \includegraphics[width=0.9\textwidth]{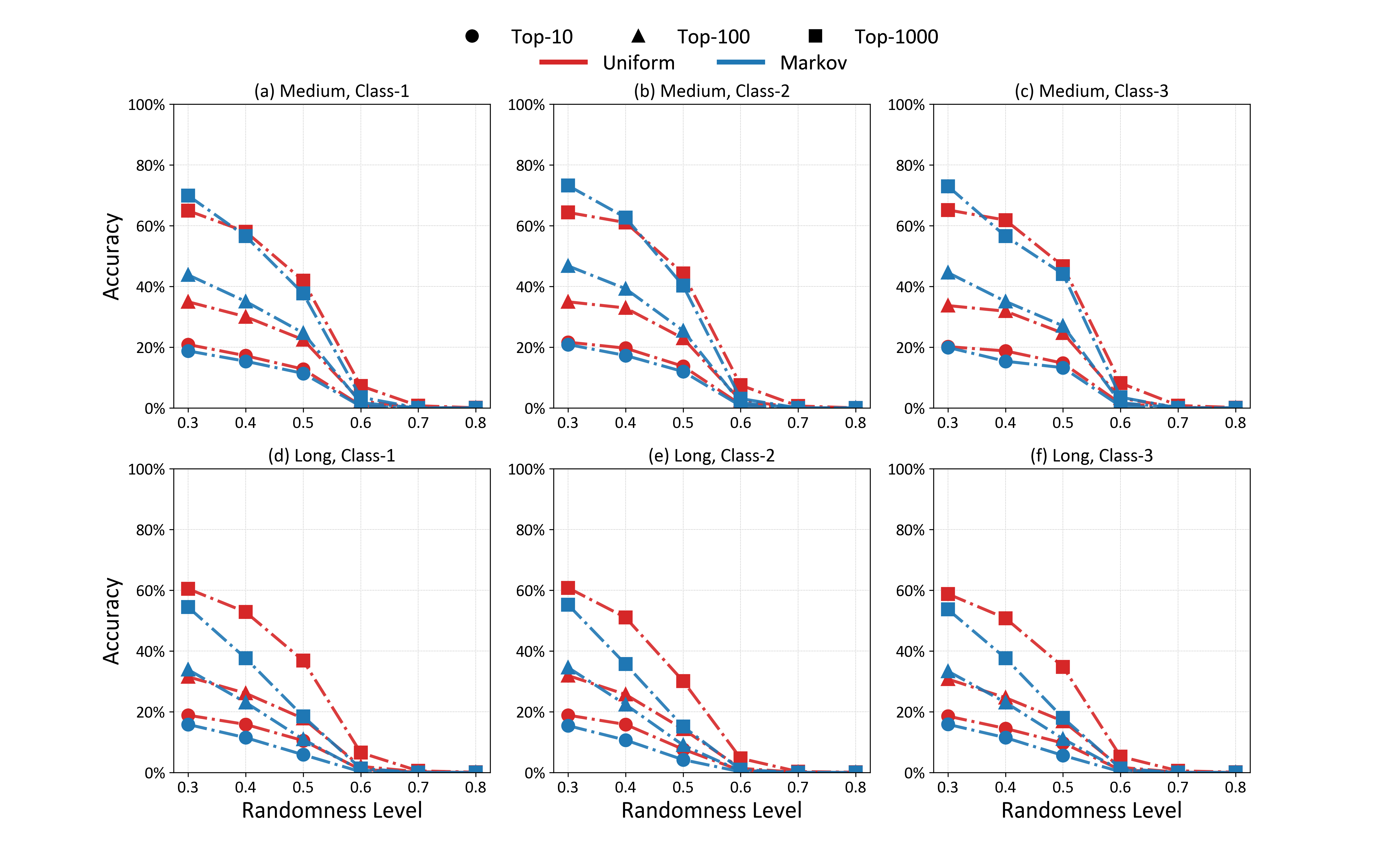} 
  \caption{Password guessing accuracy for passwords in 6 categories, for each contains curves under 2 different ghost password selection strategies and 3 different allowed guessing numbers.}
  \Description{Password guessing accuracy under different password categories.}
  \label{fig:password2x3}
  
\end{figure*}

Randomness level \(r\) also plays a critical role in affecting accuracy. When \(r\) is low\((r < 0.3)\), the algorithm strongly prefers to copy the character from the original password, resulting in a small or even minimum number of injected ghost characters, therefore, the accuracy remains high and the original password is still likely to be guessed. On the other hand, a very high randomness level \((r > 0.7)\), drives the algorithm to produce extremely long and random passwords, negatively impacting usability. From the perspective of balancing usability and security, we conclude that the ghost injection algorithm achieves a good performance under \(r \in [0.4,0.6]\), and we will discuss how to further reduce overhead with acceptable trade-off on accuracy in the following experiments.

\begin{table}[h]
\small
\centering
\caption{Guessing accuracy of Class-2 passwords under 1,000 guesses.}
\label{tab:augmentation_accuracy}
\setlength{\tabcolsep}{6pt} 
\begin{tabular}{cccc}
\hline
\textbf{Randomness} & \textbf{Length} & \multicolumn{2}{c}{\textbf{Accuracy (\%)}} \\
\cline{3-4}
\textbf{Level} & & \textbf{Uniform} & \textbf{Markov} \\
\hline
\multirow{2}{*}{0.3} 
    
    & Medium  & \textbf{64.36} & 73.22 \\
    & Long    & 60.80 & \textbf{55.28} \\
\hline
\multirow{2}{*}{0.5} 
    
    & Medium  & 44.32 & \textbf{40.28} \\
    & Long    & 30.12 & \textbf{15.14} \\
\hline
\multirow{2}{*}{0.7} 
    
    & Medium  & 0.64 & \textbf{0.00} \\
    & Long    & 0.30 & \textbf{0.04} \\
\hline
\end{tabular}

\end{table}

\noindent\textbf{\#3: Overhead in Typing.} While increasing the number of ghost characters consistently reduces the attack accuracy, it also introduces additional overhead which can potentially impact usability. As shown in Table~\ref{acc_cost}, overhead measured in moving distance always exhibits a trade-off relationship with guessing accuracy. For instance, when the randomness level is fixed at 0.4, increasing \(\lambda\) from 0 (i.e. no constraint) to 0.2 and 0.5, the overhead reduces from 34.5\% to 24.2\% and 18.1\%, respectively, but simultaneously increases adversary's guessing accuracy to 1.12\% and 3.5\%, respectively. We also see that, the main factor affecting accuracy is the randomness level, and applying constraints does not cause guessing accuracy to deviate significantly from the baseline accuracy at the same randomness level without constraints. Moreover, as more ghost characters are injected, our method achieves greater overhead reduction, thereby improving usability. These results demonstrate that \textit{VRSafe} can be flexibly configured in different methods and parameters for different user and application needs.

\noindent\textbf{\#4: Detecting Malicious Logins.} Previous experiments show that injecting ghost characters cannot prevent the adversary from guessing the original password if we allow a large number of guesses. In this experiment, we build a malicious login detector with bloom filter with maximum expected element, \(n = 10^6\) elements, expected \(FPR = 10^{-30}\) and SHA-256 as hash function to ensure enough capacity to store honeywords with low risk of false positive. We believe this is a reasonable setup for most websites that enforce a mandatory password renewal policy not exceeding one year. Given that a typical user is likely to log in thousands of times in this period~\cite{florencio2007large}, the chosen configuration ensures that a sufficient number of honeywords can be generated without exceeding the bloom filter’s maximum expected capacity. We generate a total of 20 honeywords including the ghost password using the same password guessing framework, but with only 1/10 of the data size, which is already sufficient for detecting malicious login at high accuracy. As shown in Table~\ref{tab:attempt_accuracy}, on average, 52.04\% of malicious attempts can be detected in the first guess, and 83.97\% are identified within 10 malicious login attempts. If we expand the number of honeywords to 100, the detector can achieve 72.43\% and 96.05\% detection rate under 1 and 10 login attempts, respectively.

\begin{table}[ht]
\small
\centering
\caption{Malicious login attempts detection rate under \# of honeywords = 20.}
\label{tab:attempt_accuracy}
\setlength{\tabcolsep}{6pt}

\begin{tabular}{ccc}
\hline
\textbf{Randomness} & \textbf{Login} & \textbf{Detection} \\
\textbf{Level}      & \textbf{Attempts} & \textbf{Rate (\%)} \\
\hline
\multirow{2}{*}{0.3}
    & 1   & 57.18 \\
    & 10  & 86.32 \\
\hline
\multirow{2}{*}{0.5}
    & 1   & 54.76 \\
    & 10  & 85.36 \\
\hline
\multirow{2}{*}{0.7}
    & 1   & 42.20 \\
    & 10  & 78.04 \\
\hline
\vspace{-25pt}
\end{tabular}
\end{table}

\begin{table*}[!t]
\centering
\caption{Overhead and accuracy of ghost passwords under different constraints.}
\label{acc_cost}
\newsavebox{\mytable}
\savebox{\mytable}{%
\setlength{\tabcolsep}{4pt}
\begin{tabular}{ccccccc}
\hline
\textbf{Randomness} & \textbf{Length} &
\textbf{No Constr.} &
\multicolumn{2}{c}{\textbf{Soft (Distance/Acc(\%))}} &
\multicolumn{2}{c}{\textbf{Hard (Distance/Acc(\%))}} \\
\cline{4-5} \cline{6-7}
\textbf{Level} & & Distance/Acc(\%) & $\lambda=0.2$ & $\lambda=0.5$ & $\tau=3$ & $\tau=6$ \\
\hline
\multirow{2}{*}{0.3}
    & Medium & 4.52/73.22 & 4.26/74.72 & 4.11/75.56 & 4.07/73.84 & 4.2/74.92 \\
    & Long   & 5.14/55.28 & 4.89/57.26 & 4.75/57.54 & 4.71/57.46 & 4.84/56.60 \\
\hline
\multirow{2}{*}{0.4}
    & Medium & 4.83/62.68 & 4.46/63.80 & 4.24/66.18 & 4.19/64.52 & 4.37/64.64 \\
    & Long   & 5.83/35.68 & 5.34/35.72 & 5.05/36.82 & 4.97/36.60 & 5.21/36.38 \\
\hline
\multirow{2}{*}{0.5}
    & Medium & 5.88/40.28 & 5.09/41.70 & 4.65/42.00 & 4.56/42.06 & 4.91/41.54 \\
    & Long   & 7.36/15.14 & 6.23/15.34 & 5.63/15.96 & 5.54/16.34 & 6.01/14.90 \\
\hline
\multirow{2}{*}{0.6}
    & Medium & 8.48/3.08  & 6.64/4.50  & 5.69/4.02  & 5.51/3.62  & 6.33/3.68 \\
    & Long   & 10.06/0.94 & 7.9/0.94   & 6.73/0.90  & 6.48/1.00  & 7.48/1.28 \\
\hline
\multirow{2}{*}{0.7}
    & Medium & 10.30/0.00 & 7.86/0.04  & 6.47/0.10  & 6.26/0.02  & 7.46/0.06 \\
    & Long   & 11.53/0.04 & 8.82/0.00  & 7.36/0.02  & 7.07/0.02  & 8.39/0.02 \\
\hline
\multirow{2}{*}{0.8}
    & Medium & 10.37/0.00 & 7.94/0.00  & 6.57/0.00  & 6.31/0.00  & 7.54/0.00 \\
    & Long   & 11.69/0.00 & 8.94/0.00  & 7.45/0.00  & 7.17/0.00  & 8.5/0.00  \\
\hline
\end{tabular}
}
\begin{minipage}{\wd\mytable}
    \centering
    \usebox{\mytable}
    \raggedright
    \footnotesize
    Note: Distance for original medium length passwords is 3.59, for long passwords is 4.3.
\end{minipage}
\end{table*}

{\noindent\textbf{\#5: Computing Resource Cost.} We evaluated the computing resource usage for VRSafe. The time and memory costs of the malicious login detection bloom filter are summarized in Table~\ref{tab:bf}, with a storage space cost of approximately 17.55 MB. We further measured the CPU and memory cost for running VRSafe using Android Debug Bridge (ADB), a widely used development toolkit recommended in official Meta documentation~\cite{ADB}. Over a 10-minute profiling session on Meta Quest 2, VRSafe with ghost character added during typing has a CPU utilization averaging 48\% (with a range of 35-62\%) and memory averaging \num{1.77}~\si{\giga\byte}. In comparison, typing without the ghost character (i.e. using the regular QWERTY keyboard) showed a similar average CPU utilization of 48\% (with a range of 35-52\%) and a slightly lower average memory usage of \num{1.75}~\si{\giga\byte}. These results indicate that enabling the ghost character introduces only marginal memory overhead and no noticeable increase in average CPU utilization.

\begin{table}[!h]
\small
\caption{Bloom filter time and space cost.}
\label{tab:bf}
\centering
\setlength{\tabcolsep}{6pt}
\begin{tabular}{ccc}
\hline
\textbf{Operation} & \textbf{Time (s)} & \textbf{Memory (KiB)} \\
\hline
Init   & 4.20e-05 & 0.37 \\
Insert & 7.02e-06 & 4.38 \\
Lookup & 4.35e-06 & 4.38 \\
\hline
\end{tabular}
\end{table}

\subsection{User Study}\label{userstudy}

\noindent\textbf{Methods.} Our study is approved by our Institutional Review Board (IRB). To simulate user account logins using password within a VR environment, we developed a VR application using Unity 2022.3.24f1 that uses \textit{VRSafe} as the keyboard for typing (Fig.~\ref{fig:VRSafeInference}). Participants in our study were seated in a chair and instructed to complete the password typing tasks, and their actions were recorded by a camera to simulate video-based keystroke inference attacks (see Fig.~\ref{fig:Camera} in Appendix~\ref{appendix:userstudy}). All participants were informed of the recording and provided consent. To protect participants' privacy, facial information was excluded from the recordings, and the passwords used for the typing tasks were selected from the publicly available leaked datasets. We did not initially disclose that the recording simulated an attack scenario, to avoid influencing participants' natural typing behavior. After completing the study, we fully debriefed participants on the experiment design, including the purpose of the video recording, and addressed any concerns or questions. Each participant received 10 USD as appreciation. Participants were first introduced to the essential knowledge for using the VR device (e.g., moving the cursor, typing), as well as the differences between a standard keyboard and the \textit{VRSafe} keyboard. They were then given five minutes to explore the application freely. The main experiment consisted of ten password typing tasks. The first two tasks involved standard keyboard input without ghost characters, serving as a baseline. The remaining eight tasks incorporated ghost characters, simulating the enhanced security mechanism of \textit{VRSafe}. After completing each task, the participant notifies the researcher and clicks the “Next” button on the login page to proceed to the next task. Upon completion of all tasks, participants were asked to complete a post-study questionnaire regarding their experience. We used MediaPipe~\cite{zhang2020mediapipe}, a popular hand-tracking framework to extract the keystrokes from the collected video clips.

\noindent {\bf Results:} We recruited 15 participants (mean age = 25.6, sd = 4.22; 7 males, 8 females) in this study through printed flyers and announcements in weekly newsletters. All participants were over 18 years old. There were no restrictions regarding student status or prior VR experience, and the only requirement was for the participant to have a normal or corrected-to-normal vision when wearing the VR headset. Among the 15 participants, 9 do not have any prior experience with VR headset, and the remaining 6 use VR devices less than once per month. A total of 150 video clips were recorded, of which 6 were discarded.~\footnote{Mediapipe failed to recognize hands in 2 videos since the participant had long wearable nail extensions and 3 videos were discarded since the participant misunderstood the ghost character typing process as a bug and terminated the task. 1 video was discarded because the hand tracking was lost and the participant could not proceed with the task.} The average password length across all tasks was 8.35. The average number of keystrokes detected in the baseline tasks (without ghost characters) was 10.5, while the average in the remaining eight tasks (with ghost characters) was 16.2.

We find that the "false keystrokes" are treated as authentic ones by the threat model and injected to the inferred keystroke inference as expected. We examine if using \textit{VRSafe} will result in longer entry time compared to a normal QWERTY keyboard. In Fig.~\ref{fig: Password Entry Time}, we plot the entry times of all 15 participants across the 10 typing tasks and we highlight the first two tasks (i.e. no ghost characters) with red dots. We observe that the average entry time of most participants fall within 40 to 60 seconds, with no evident pattern suggesting that tasks with added ghost characters take significantly longer. For instance, some participants (e.g. P2, P13) completed baseline tasks more quickly, while others (e.g. P1, P8) showed longer or mixed durations. As the visual distributions do not suggest a clear difference in entry time between the conditions, we aggregated data across all participants and applied the Mann-Whitney U test~\cite{mann1947test} to formally assess the effect of task typing when ghost characters are included. The analysis indicates a statistically significant difference between the two groups \((U = 1148.5, p = 0.0072 < 0.05)\). To further investigate the source of the observed difference, we perform pair-wise tests. Specifically, we treated the two baseline tasks from each of the 15 participants as a single baseline group (totaling 2×15=30 samples), and the remaining tasks were grouped by their task index (i.e. all third tasks from each participant, all fourth tasks, and so on). We then apply pair-wise tests between the baseline group and the other groups one by one, and the third task group (i.e. first task involving ghost characters) shows the greatest difference from the baseline, suggesting it is the primary driver of the overall significant difference. It indicates that even though participants were already aware of the ghost character mechanism, they still paused when they first encountered it. We attribute this to a natural learning curve, and expect the impact to diminish as users become more familiar with the application.

\begin{figure*}[h]
    \centering
    \includegraphics[width=0.7\textwidth]{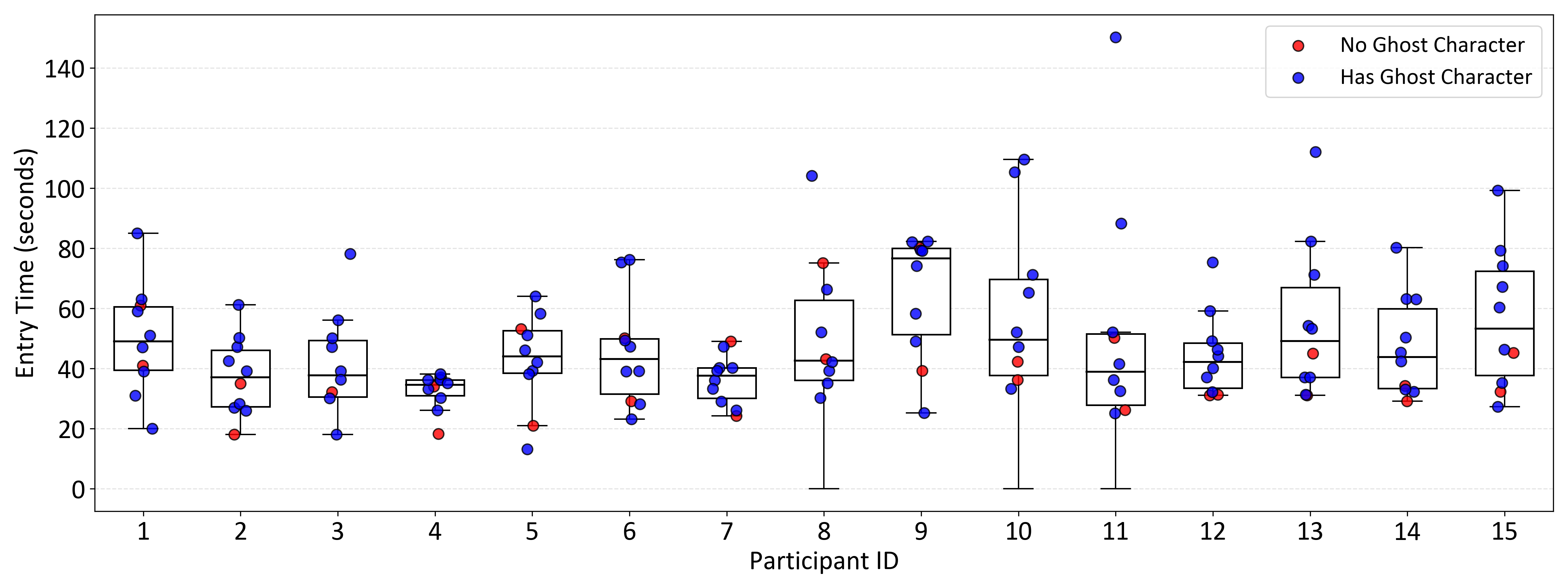}
    \caption{Password entry time of all participants in all 10 tasks.}
    \Description{Password entry time of all participants.}
    \label{fig: Password Entry Time}
\end{figure*}

In the post-task questionnaire, we use the System Usability Scale (SUS)~\cite{brooke1996sus} to evaluate Perceived Ease of Use (PEOU), Perceived Usefulness (PU) and Behavioral Intention to Use (BI) from the Technology Acceptance Model (TAM)~\cite{4af0105e-77c6-3217-8650-3ef6ef5e917a} to measure the user's acceptance.~\footnote{SUS score ranges from 0-100; PEOU, PU and BI range from 0-5.} The full questionnaire can be found in Table~\ref{tab:vr_keyboard_survey} (Appendix~\ref{appendix:userstudy}). We also test the validity and reliability of our question setting using Cronbach's Alpha~\cite{cronbach1951coefficient} and Average Variance Extracted (AVE)~\cite{fornell1981evaluating} (Table~\ref{tab:Cronbach}, Appendix~\ref{appendix:userstudy}). All Cronbach's Alpha scores are around or above 0.8, and AVEs exceed 0.6, indicating that our questions effectively evaluate the design across different aspects in the same direction. The statistical results are shown in Table~\ref{tab:post_task_feedback}, and per-question descriptive statistics of SUS are provided in Table~\ref{tab:sus_questions}(Appendix~\ref{appendix:userstudy}). We also ask some open-ended questions to encourage participants to share their thoughts freely, and we find that participants evaluate \textit{VRSafe} differently depending on the aspects they prioritize. Some participants appreciated the enhanced security and expressed a willingness to adopt even more sophisticated methods. Others thought that typing in VR is already challenging, adding extra steps could become overwhelming even if it becomes more secure. This divergence in user perspectives helps explain the wide variance observed in the SUS scores. Participants also comment on general VR device usage experience such as "improve responsiveness", "control can be more accurate", "headset feels too heavy" which are not directly related to the \textit{VRSafe} design.

\begin{table}[h]
\centering
\caption{Statistical results of participants' survey feedback.}
\label{tab:post_task_feedback}
\setlength{\tabcolsep}{6pt}
\begin{tabular}{lcccc}
\hline
\textbf{Item} & \textit{Min} & \textit{Max} & \textit{Mean} & \textit{Sd} \\
\hline
PEOU & 1.75 & 4.75 & 3.88 & 0.83 \\
PU   & 2.33 & 5.00 & 3.38 & 0.80 \\
BI   & 1.50 & 5.00 & 3.30 & 1.28 \\
SUS  & 25   & 90   & 58.6 & 14.4 \\
\hline
\end{tabular}
\end{table}

\section{Conclusions}\label{futurework}

In this paper, we present \textit{VRSafe}, a VR keyboard to protect users from keystroke inference attacks. The proposed solution is designed to enhance the word-level security of password inputs for VR users using a standard virtual QWERTY keyboard layout. \textit{VRSafe} is the first novel solution to incorporate a detection mechanism to identify unauthorized login attempts using credentials inferred from keystroke inference attacks. Our experiment evaluation shows that \textit{VRSafe} is effective against attacks based on password guessing models and incurs only a modest overhead to achieve enhanced password security and detect malicious login attempts. The results of our user study indicate that \textit{VRSafe} can be a highly usable alternative for normal users with higher security needs. \textit{VRSafe} is also potentially applicable to other forms of keystroke inference attacks including acoustic-based and Wifi-based attacks and our future work is focused on extending \textit{VRSafe} to protect against those forms of attacks. We also plan to investigate mechanisms for protecting sensitive text entered by virtual avatars in extended reality applications. Unlike conventional keystroke attacks, where adversaries observe a victim’s physical actions in the real world, extended reality environments allow adversaries to directly inspect the actions of virtual avatars, potentially revealing additional information that can be exploited to infer keystrokes.

\begin{acks}

This material is based upon work supported by the National Science Foundation under Grant \#2211507. Any opinions,
findings, and conclusions or recommendations expressed in this material are those of the authors and do not necessarily
reflect the views of the National Science Foundation. The authors also acknowledge the partial support for this work through a grant from the Institute for Cyber Law, Policy, and Security (Pitt Cyber) at the University of Pittsburgh.
\end{acks}

\bibliographystyle{ACM-Reference-Format}
\balance
\bibliography{sample-base}

@inproceedings{luo2022holologger,
  title={Holologger: Keystroke inference on mixed reality head mounted displays},
  author={Luo, Shiqing and Hu, Xinyu and Yan, Zhisheng},
  booktitle={2022 IEEE Conference on Virtual Reality and 3D User Interfaces (VR)},
  pages={445--454},
  year={2022},
  organization={IEEE}
}

@inproceedings{meteriz2022keylogging,
  title={A keylogging inference attack on air-tapping keyboards in virtual environments},
  author={Meteriz-Y{\i}ld{\i}ran, {\"U}lk{\"u} and Y{\i}ld{\i}ran, Necip Faz{\i}l and Awad, Amro and Mohaisen, David},
  booktitle={2022 IEEE Conference on Virtual Reality and 3D User Interfaces (VR)},
  pages={765--774},
  year={2022},
  organization={IEEE}
}

@inproceedings{al2021vr,
  title={Vr-spy: A side-channel attack on virtual key-logging in vr headsets},
  author={Al Arafat, Abdullah and Guo, Zhishan and Awad, Amro},
  booktitle={2021 IEEE Virtual Reality and 3D User Interfaces (VR)},
  pages={564--572},
  year={2021},
  organization={IEEE}
}

@article{yang2022wireless,
  title={Wireless training-free keystroke inference attack and defense},
  author={Yang, Edwin and Fang, Song and Markwood, Ian and Liu, Yao and Zhao, Shangqing and Lu, Zhuo and Zhu, Haojin},
  journal={IEEE/ACM Transactions on Networking},
  volume={30},
  number={4},
  pages={1733--1748},
  year={2022},
  publisher={IEEE}
}

@inproceedings{sun2016visible,
  title={Visible: Video-assisted keystroke inference from tablet backside motion.},
  author={Sun, Jingchao and Jin, Xiaocong and Chen, Yimin and Zhang, Jinxue and Zhang, Yanchao and Zhang, Rui},
  booktitle={NDSS},
  year={2016}
}

@inproceedings{ling2019know,
  title={I know what you enter on gear vr},
  author={Ling, Zhen and Li, Zupei and Chen, Chen and Luo, Junzhou and Yu, Wei and Fu, Xinwen},
  booktitle={2019 IEEE Conference on Communications and Network Security (CNS)},
  pages={241--249},
  year={2019},
  organization={IEEE}
}

@inproceedings{luo2024eavesdropping,
  title={Eavesdropping on controller acoustic emanation for keystroke inference attack in virtual reality},
  author={Luo, Shiqing and Nguyen, Anh and Farooq, Hafsa and Sun, Kun and Yan, Zhisheng},
  booktitle={The Network and Distributed System Security Symposium (NDSS)},
  year={2024}
}

@article{lee2023vrkeylogger,
  title={VRKeyLogger: Virtual keystroke inference attack via eavesdropping controller usage pattern in WebVR},
  author={Lee, Jiyeon and Kim, Hyosu and Lee, Kilho},
  journal={Computers \& Security},
  volume={134},
  pages={103461},
  year={2023},
  publisher={Elsevier}
}

@inproceedings{yang2023towards,
  title={Towards a general video-based keystroke inference attack},
  author={Yang, Zhuolin and Chen, Yuxin and Sarwar, Zain and Schwartz, Hadleigh and Zhao, Ben Y and Zheng, Haitao},
  booktitle={32nd USENIX Security Symposium (USENIX Security 23)},
  pages={141--158},
  year={2023}
}

@inproceedings{gopal2023hidden,
  title={Hidden reality: Caution, your hand gesture inputs in the immersive virtual world are visible to all!},
  author={Gopal, Sindhu Reddy Kalathur and Shukla, Diksha and Wheelock, James David and Saxena, Nitesh},
  booktitle={32nd USENIX Security Symposium (USENIX Security 23)},
  pages={859--876},
  year={2023}
}

@article{zhang2020mediapipe,
  title={Mediapipe hands: On-device real-time hand tracking},
  author={Zhang, Fan and Bazarevsky, Valentin and Vakunov, Andrey and Tkachenka, Andrei and Sung, George and Chang, Chuo-Ling and Grundmann, Matthias},
  journal={arXiv preprint arXiv:2006.10214},
  year={2020}
}

@inproceedings{pal2019beyond,
  title={Beyond credential stuffing: Password similarity models using neural networks},
  author={Pal, Bijeeta and Daniel, Tal and Chatterjee, Rahul and Ristenpart, Thomas},
  booktitle={2019 IEEE Symposium on Security and Privacy (SP)},
  pages={417--434},
  year={2019},
  organization={IEEE}
}

@inproceedings{wang2023pass2edit,
  title={$\{$Pass2Edit$\}$: A $\{$Multi-Step$\}$ Generative Model for Guessing Edited Passwords},
  author={Wang, Ding and Zou, Yunkai and Xiao, Yuan-An and Ma, Siqi and Chen, Xiaofeng},
  booktitle={32nd USENIX Security Symposium (USENIX Security 23)},
  pages={983--1000},
  year={2023}
}

@article{wan2024analysis,
  title={Analysis and Design of Efficient Authentication Techniques for Password Entry with the Qwerty Keyboard for VR Environments},
  author={Wan, Tingjie and Zhang, Liangyuting and Xu, Yunxin and Guo, Zixuan and Gao, Boyu and Liang, Hai-Ning},
  journal={IEEE Transactions on Visualization and Computer Graphics},
  year={2024},
  publisher={IEEE}
}

@inproceedings{ma2014study,
  title={A study of probabilistic password models},
  author={Ma, Jerry and Yang, Weining and Luo, Min and Li, Ninghui},
  booktitle={2014 IEEE Symposium on Security and Privacy},
  pages={689--704},
  year={2014},
  organization={IEEE}
}

@misc{ADB,
  author =       {Meta},
  year         = {2024},
  url          = {https://developers.meta.com/horizon/documentation/native/android/ts-adb/},
  note         = {Accessed: 2024-10-30}
}

@misc{MRTK,
  author =       {Microsoft},
  year         = {2024},
  url          = {https://github.com/microsoft/MixedRealityToolkit-Unity},
  note         = {\url{https://github.com/microsoft/MixedRealityToolkit-Unity}}
}

@article{boutros2020iris,
  title={Iris and periocular biometrics for head mounted displays: Segmentation, recognition, and synthetic data generation},
  author={Boutros, Fadi and Damer, Naser and Raja, Kiran and Ramachandra, Raghavendra and Kirchbuchner, Florian and Kuijper, Arjan},
  journal={Image and Vision Computing},
  volume={104},
  pages={104007},
  year={2020},
  publisher={Elsevier}
}

@inproceedings{luo2020oculock,
  title={OcuLock: Exploring human visual system for authentication in virtual reality head-mounted display},
  author={Luo, Shiqing and Nguyen, Anh and Song, Chen and Lin, Feng and Xu, Wenyao and Yan, Zhisheng},
  booktitle={2020 Network and Distributed System Security Symposium (NDSS)},
  year={2020}
}

@inproceedings{wang2016targeted,
  title={Targeted online password guessing: An underestimated threat},
  author={Wang, Ding and Zhang, Zijian and Wang, Ping and Yan, Jeff and Huang, Xinyi},
  booktitle={Proceedings of the 2016 ACM SIGSAC conference on computer and communications security},
  pages={1242--1254},
  year={2016}
}

@inproceedings{maiti2017randompad,
  title={Randompad: Usability of randomized mobile keypads for defeating inference attacks},
  author={Maiti, Anindya and Crager, Kirsten},
  booktitle={Proceedings of the IEEE Euro S\&P Workshop on Innovations in Mobile Privacy \& Security (IMPS)},
  year={2017}
}

@InProceedings{10.1007/978-981-99-8024-6_13,
author="Althebeiti, Hattan
and Gedawy, Ran
and Alghuried, Ahod
and Nyang, Daehun
and Mohaisen, David",
editor="Kim, Howon
and Youn, Jonghee",
title="Defending AirType Against Inference Attacks Using 3D In-Air Keyboard Layouts: Design and Evaluation",
booktitle="Information Security Applications",
year="2024",
publisher="Springer Nature Singapore",
address="Singapore",
pages="159--174",
isbn="978-981-99-8024-6"
}

@article{wang2024muki,
  title={MuKI-Fi: Multi-person keystroke inference with BFI-enabled Wi-Fi sensing},
  author={Wang, Hongbo and Hu, Jingyang and Zheng, Tianyue and Hu, Jingzhi and Chen, Zhe and Jiang, Hongbo and Zheng, Yuanjin and Luo, Jun},
  journal={IEEE Transactions on Mobile Computing},
  year={2024},
  publisher={IEEE}
}

@inproceedings{zhang2023s,
  title={It's all in your head (set): Side-channel attacks on $\{$AR/VR$\}$ systems},
  author={Zhang, Yicheng and Slocum, Carter and Chen, Jiasi and Abu-Ghazaleh, Nael},
  booktitle={32nd USENIX Security Symposium (USENIX Security 23)},
  pages={3979--3996},
  year={2023}
}

@inproceedings{slocum2023going,
  title={Going through the motions:$\{$AR/VR$\}$ keylogging from user head motions},
  author={Slocum, Carter and Zhang, Yicheng and Abu-Ghazaleh, Nael and Chen, Jiasi},
  booktitle={32nd USENIX Security Symposium (USENIX Security 23)},
  pages={159--174},
  year={2023}
}

@inproceedings{wu2023privacy,
  title={Privacy leakage via unrestricted motion-position sensors in the age of virtual reality: A study of snooping typed input on virtual keyboards},
  author={Wu, Yi and Shi, Cong and Zhang, Tianfang and Walker, Payton and Liu, Jian and Saxena, Nitesh and Chen, Yingying},
  booktitle={2023 IEEE Symposium on Security and Privacy (SP)},
  pages={3382--3398},
  year={2023},
  organization={IEEE}
}

@article{sabra2020zoom,
  title={Zoom on the keystrokes: Exploiting video calls for keystroke inference attacks},
  author={Sabra, Mohd and Maiti, Anindya and Jadliwala, Murtuza},
  journal={arXiv preprint arXiv:2010.12078},
  year={2020}
}

@article{ni2024non,
  title={Non-intrusive and Unconstrained Keystroke Inference in VR Platforms via Infrared Side Channel},
  author={Ni, Tao and Du, Yuefeng and Zhao, Qingchuan and Wang, Cong},
  journal={arXiv preprint arXiv:2412.14815},
  year={2024}
}

@inproceedings{wang2024gazeploit,
  title={GAZEploit: Remote Keystroke Inference Attack by Gaze Estimation from Avatar Views in VR/MR Devices},
  author={Wang, Hanqiu and Zhan, Zihao and Shan, Haoqi and Dai, Siqi and Panoff, Maximilian and Wang, Shuo},
  booktitle={Proceedings of the 2024 on ACM SIGSAC Conference on Computer and Communications Security},
  pages={1731--1745},
  year={2024}
}

@article{brooke1996sus,
  title={SUS-A quick and dirty usability scale},
  author={Brooke, John and others},
  journal={Usability evaluation in industry},
  volume={189},
  number={194},
  pages={4--7},
  year={1996},
  publisher={London, England.}
}

@article{4af0105e-77c6-3217-8650-3ef6ef5e917a,
 ISSN = {02767783, 21629730},
 URL = {http://www.jstor.org/stable/249008},
 author = {Fred D. Davis},
 journal = {MIS Quarterly},
 number = {3},
 pages = {319--340},
 publisher = {Management Information Systems Research Center, University of Minnesota},
 title = {Perceived Usefulness, Perceived Ease of Use, and User Acceptance of Information Technology},
 urldate = {2025-03-09},
 volume = {13},
 year = {1989}
}

@misc{Beyond_reality,
  author       = {National Research Group},
  title        = {Beyond reality: Is the VR revolution on the horizon?},
  year         = {2022},
  howpublished = {\url{https://www.nrgmr.com/our-thinking/technology/the-vr-revolution-might-finally-be-on-the-horizon/}},
  note         = {Accessed: 2025-04-10}
}

@inproceedings{stephenson2022sok,
  title={Sok: Authentication in augmented and virtual reality},
  author={Stephenson, Sophie and Pal, Bijeeta and Fan, Stephen and Fernandes, Earlence and Zhao, Yuhang and Chatterjee, Rahul},
  booktitle={2022 IEEE symposium on security and privacy (SP)},
  pages={267--284},
  year={2022},
  organization={IEEE}
}

@inproceedings{maiti2017preventing,
  title={Preventing shoulder surfing using randomized augmented reality keyboards},
  author={Maiti, Anindya and Jadliwala, Murtuza and Weber, Chase},
  booktitle={2017 IEEE international conference on pervasive computing and communications workshops (PerCom Workshops)},
  pages={630--635},
  year={2017},
  organization={IEEE}
}

@inproceedings{DBLP:conf/ccs/JuelsR13,
  author       = {Ari Juels and
                  Ronald L. Rivest},
  editor       = {Ahmad{-}Reza Sadeghi and
                  Virgil D. Gligor and
                  Moti Yung},
  title        = {Honeywords: making password-cracking detectable},
  booktitle    = {2013 {ACM} {SIGSAC} Conference on Computer and Communications Security,
                  CCS'13, Berlin, Germany, November 4-8, 2013},
  pages        = {145--160},
  publisher    = {{ACM}},
  year         = {2013},
  url          = {https://doi.org/10.1145/2508859.2516671},
  doi          = {10.1145/2508859.2516671},
  bibsource    = {dblp computer science bibliography, https://dblp.org}
}

@inproceedings{DBLP:conf/ndss/WangR24,
  author       = {Ke Coby Wang and
                  Michael K. Reiter},
  title        = {Bernoulli Honeywords},
  booktitle    = {31st Annual Network and Distributed System Security Symposium, {NDSS}
                  2024, San Diego, California, USA, February 26 - March 1, 2024},
  publisher    = {The Internet Society},
  year         = {2024},
  url          = {https://www.ndss-symposium.org/ndss-paper/bernoulli-honeywords/},
  bibsource    = {dblp computer science bibliography, https://dblp.org}
}

@inproceedings{DBLP:conf/uss/HuangBR24,
  author       = {Zonghao Huang and
                  Lujo Bauer and
                  Michael K. Reiter},
  editor       = {Davide Balzarotti and
                  Wenyuan Xu},
  title        = {The Impact of Exposed Passwords on Honeyword Efficacy},
  booktitle    = {33rd {USENIX} Security Symposium, {USENIX} Security 2024, Philadelphia,
                  PA, USA, August 14-16, 2024},
  publisher    = {{USENIX} Association},
  year         = {2024},
  url          = {https://www.usenix.org/conference/usenixsecurity24/presentation/huang-zonghao},
  bibsource    = {dblp computer science bibliography, https://dblp.org}
}

@inproceedings{DBLP:conf/sp/0002ZDSH22,
  author       = {Ding Wang and
                  Yunkai Zou and
                  Qiying Dong and
                  Yuanming Song and
                  Xinyi Huang},
  title        = {How to Attack and Generate Honeywords},
  booktitle    = {43rd {IEEE} Symposium on Security and Privacy, {SP} 2022, San Francisco,
                  CA, USA, May 22-26, 2022},
  pages        = {966--983},
  publisher    = {{IEEE}},
  year         = {2022},
  url          = {https://doi.org/10.1109/SP46214.2022.9833598},
  doi          = {10.1109/SP46214.2022.9833598},
  bibsource    = {dblp computer science bibliography, https://dblp.org}
}

@article{bloom1970space,
  title={Space/time trade-offs in hash coding with allowable errors},
  author={Bloom, Burton H},
  journal={Communications of the ACM},
  volume={13},
  number={7},
  pages={422--426},
  year={1970},
  publisher={ACM New York, NY, USA}
}

@misc{VRStatReport2025,
  author       = {Naveen Kumar},
  title        = {Virtual Reality Statistics 2025: Users \& Trends},
  howpublished = {\url{https://www.demandsage.com/virtual-reality-statistics/}},
  year         = {2025},
  note         = {Accessed: 2025-07-26}
}

@article{solso1979bigram,
  title={Bigram and trigram frequencies and versatilities in the English language},
  author={Solso, Robert L and Barbuto, Paul F and Juel, Connie L},
  journal={Behavior Research Methods \& Instrumentation},
  volume={11},
  number={5},
  pages={475--484},
  year={1979},
  publisher={Springer}
}

@inproceedings{florencio2007large,
  title={A large-scale study of web password habits},
  author={Florencio, Dinei and Herley, Cormac},
  booktitle={Proceedings of the 16th international conference on World Wide Web},
  pages={657--666},
  year={2007}
}

@misc{COMB_news,
  author =       "Ron Cresswell, J.D., CFE",
  year =         "2021",
  title =        "The COMB Data Leak: What You Should Know",
  url =          "https://www.acfe.com/acfe-insights-blog/blog-detail?s=comb-data-leak-what-you-should-know",
  lastaccessed = "August 28, 2025",
}

@inproceedings{weir2009password,
  title={Password cracking using probabilistic context-free grammars},
  author={Weir, Matt and Aggarwal, Sudhir and De Medeiros, Breno and Glodek, Bill},
  booktitle={2009 30th IEEE symposium on security and privacy},
  pages={391--405},
  year={2009},
  organization={IEEE}
}

@article{mann1947test,
  title={On a test of whether one of two random variables is stochastically larger than the other},
  author={Mann, Henry B and Whitney, Donald R},
  journal={The annals of mathematical statistics},
  pages={50--60},
  year={1947},
  publisher={JSTOR}
}

@article{vaswani2017attention,
  title={Attention is all you need},
  author={Vaswani, Ashish and Shazeer, Noam and Parmar, Niki and Uszkoreit, Jakob and Jones, Llion and Gomez, Aidan N and Kaiser, {\L}ukasz and Polosukhin, Illia},
  journal={Advances in neural information processing systems},
  volume={30},
  year={2017}
}

@article{sutskever2014sequence,
  title={Sequence to sequence learning with neural networks},
  author={Sutskever, Ilya and Vinyals, Oriol and Le, Quoc V},
  journal={Advances in neural information processing systems},
  volume={27},
  year={2014}
}

@article{achiam2023gpt,
  title={Gpt-4 technical report},
  author={Achiam, Josh and Adler, Steven and Agarwal, Sandhini and Ahmad, Lama and Akkaya, Ilge and Aleman, Florencia Leoni and Almeida, Diogo and Altenschmidt, Janko and Altman, Sam and Anadkat, Shyamal and others},
  journal={arXiv preprint arXiv:2303.08774},
  year={2023}
}

@article{cronbach1951coefficient,
  title={Coefficient alpha and the internal structure of tests},
  author={Cronbach, Lee J},
  journal={psychometrika},
  volume={16},
  number={3},
  pages={297--334},
  year={1951},
  publisher={Springer-Verlag}
}

@article{fornell1981evaluating,
  title={Evaluating structural equation models with unobservable variables and measurement error},
  author={Fornell, Claes and Larcker, David F},
  journal={Journal of marketing research},
  volume={18},
  number={1},
  pages={39--50},
  year={1981},
  publisher={Sage Publications Sage CA: Los Angeles, CA}
}

@inproceedings{melicher2016fast,
  title={Fast, lean, and accurate: Modeling password guessability using neural networks},
  author={Melicher, William and Ur, Blase and Segreti, Sean M and Komanduri, Saranga and Bauer, Lujo and Christin, Nicolas and Cranor, Lorrie Faith},
  booktitle={25th USENIX Security Symposium (USENIX Security 16)},
  pages={175--191},
  year={2016}
}

@inproceedings{kelley2012guess,
  title={Guess again (and again and again): Measuring password strength by simulating password-cracking algorithms},
  author={Kelley, Patrick Gage and Komanduri, Saranga and Mazurek, Michelle L and Shay, Richard and Vidas, Timothy and Bauer, Lujo and Christin, Nicolas and Cranor, Lorrie Faith and Lopez, Julio},
  booktitle={2012 IEEE symposium on security and privacy},
  pages={523--537},
  year={2012},
  organization={IEEE}
}

@article{nosenko2023password,
  title={Password and passphrase guessing with recurrent neural networks},
  author={Nosenko, Alex and Cheng, Yuan and Chen, Haiquan},
  journal={Information Systems Frontiers},
  volume={25},
  number={2},
  pages={549--565},
  year={2023},
  publisher={Springer}
}

@article{gupta2018side,
  title={A side-channel attack on smartphones: Deciphering key taps using built-in microphones},
  author={Gupta, Haritabh and Sural, Shamik and Atluri, Vijayalakshmi and Vaidya, Jaideep},
  journal={Journal of computer security},
  volume={26},
  number={2},
  pages={255--281},
  year={2018},
  publisher={SAGE Publications Sage UK: London, England}
}

@inproceedings{cronin2021charger,
  title={$\{$Charger-Surfing$\}$: Exploiting a power line $\{$Side-Channel$\}$ for smartphone information leakage},
  author={Cronin, Patrick and Gao, Xing and Yang, Chengmo and Wang, Haining},
  booktitle={30th USENIX Security Symposium (USENIX Security 21)},
  pages={681--698},
  year={2021}
}

@inproceedings{yang2025targeted,
  title={Targeted Password Guessing Using Neural Language Models},
  author={Yang, Jiahong and Li, Wenting and Cheng, Haibo and Wang, Ping},
  booktitle={ICASSP 2025-2025 IEEE International Conference on Acoustics, Speech and Signal Processing (ICASSP)},
  pages={1--5},
  year={2025},
  organization={IEEE}
}

@inproceedings{soni2021keynet,
  title={KeyNet: enhancing cybersecurity with deep learning-based LSTM on keystroke dynamics for authentication},
  author={Soni, Jayesh and Prabakar, Nagarajan},
  booktitle={International Conference on Intelligent Human Computer Interaction},
  pages={761--771},
  year={2021},
  organization={Springer}
}

@article{shen2018gaitlock,
  title={GaitLock: Protect virtual and augmented reality headsets using gait},
  author={Shen, Yiran and Wen, Hongkai and Luo, Chengwen and Xu, Weitao and Zhang, Tao and Hu, Wen and Rus, Daniela},
  journal={IEEE Transactions on Dependable and Secure Computing},
  volume={16},
  number={3},
  pages={484--497},
  year={2018},
  publisher={IEEE}
}

@article{li2020designing,
  title={Designing leakage-resilient password entry on head-mounted smart wearable glass devices},
  author={Li, Yan and Cheng, Yao and Meng, Weizhi and Li, Yingjiu and Deng, Robert H},
  journal={IEEE Transactions on Information Forensics and security},
  volume={16},
  pages={307--321},
  year={2020},
  publisher={IEEE}
}

@misc{Meta_locomotion,
  author =       "Meta Platforms, Inc",
  year =         "2025",
  title =        "Locomotion types",
  url =          "https://developers.meta.com/horizon/design/locomotion-types/",
  lastaccessed = "Dec 2, 2025",
}

\section*{Ethics Statement}
All data collection procedures were reviewed and approved by IRB (STUDY24120080) to ensure that no personally identifiable information or sensitive user data was gathered without explicit consent. Participants were informed of the study’s purpose, potential risks, and their right to withdraw at any time. All data was stored securely to prevent unauthorized access. The results of this work are presented solely for scientific and educational purposes and are not intended to facilitate any malicious activity.

\section*{Open Science Statement}
To support reproducibility and foster future research on VR input security, we will release our VRSafe virtual keyboard implementation in Unity along with the corresponding source code at https://github.com/odinyuan/VRSafe. For the evaluation dataset (i.e. the COMB dataset), although the leak is publicly available online, we refrain from including its URL to avoid further dissemination. Interested researchers may contact the authors for details. All evaluation code is also made publicly available in the same repository. By making these resources available, we encourage researchers to replicate our findings, evaluate alternative attack and defense mechanisms, and extend \textit{VRSafe} to broader VR interaction contexts.

\appendix

\section{User Study}
\label{appendix:userstudy}

\begin{center}
\captionof{table}{SUS Question Statistics}
\label{tab:sus_questions}
\begin{tabular}{cp{5cm}cc}
\hline
\textbf{\#} & \textbf{Question} & \textbf{Mean} & \textbf{Sd} \\
\hline
1  & I think I would like to use this VR Keyboard frequently.           & 2.73 & 1.22 \\
2  & I found the VR Keyboard unnecessarily complex.                     & 2.80 & 1.14 \\
3  & I thought the VR Keyboard was easy to use.                         & 3.46 & 1.30 \\
4  & I think that I would need the support of a technical person to be able to use the VR Keyboard.            & 2.33 & 1.34 \\
5  & I found the various functions in this VR Keyboard were well integrated.                    & 3.73 & 0.70 \\
6  & I thought there was too much inconsistency in this VR Keyboard.                 & 2.73 & 1.03 \\                                             
7  & I would imagine that most people would learn to use this VR Keyboard very quickly.               & 4.06 & 0.79 \\
8  & I found this VR Keyboard very cumbersome to use.                   & 3.06 & 1.22 \\
9  & I felt very confident using VR Keyboard.                           & 3.13 & 1.24 \\
10 & I needed to learn a lot of things before I could get going with this VR Keyboard.              & 2.73 & 1.22 \\
\hline
\multicolumn{4}{l}{\small\textit{Ratings on a 5-point Likert scale}} \\
\multicolumn{4}{l}{\small\textit{(1 = Strongly Disagree, 5 = Strongly Agree).}} \\
\end{tabular}
\end{center}

\begin{figure}[H]
  \centering
  \includegraphics[width=0.45\textwidth]{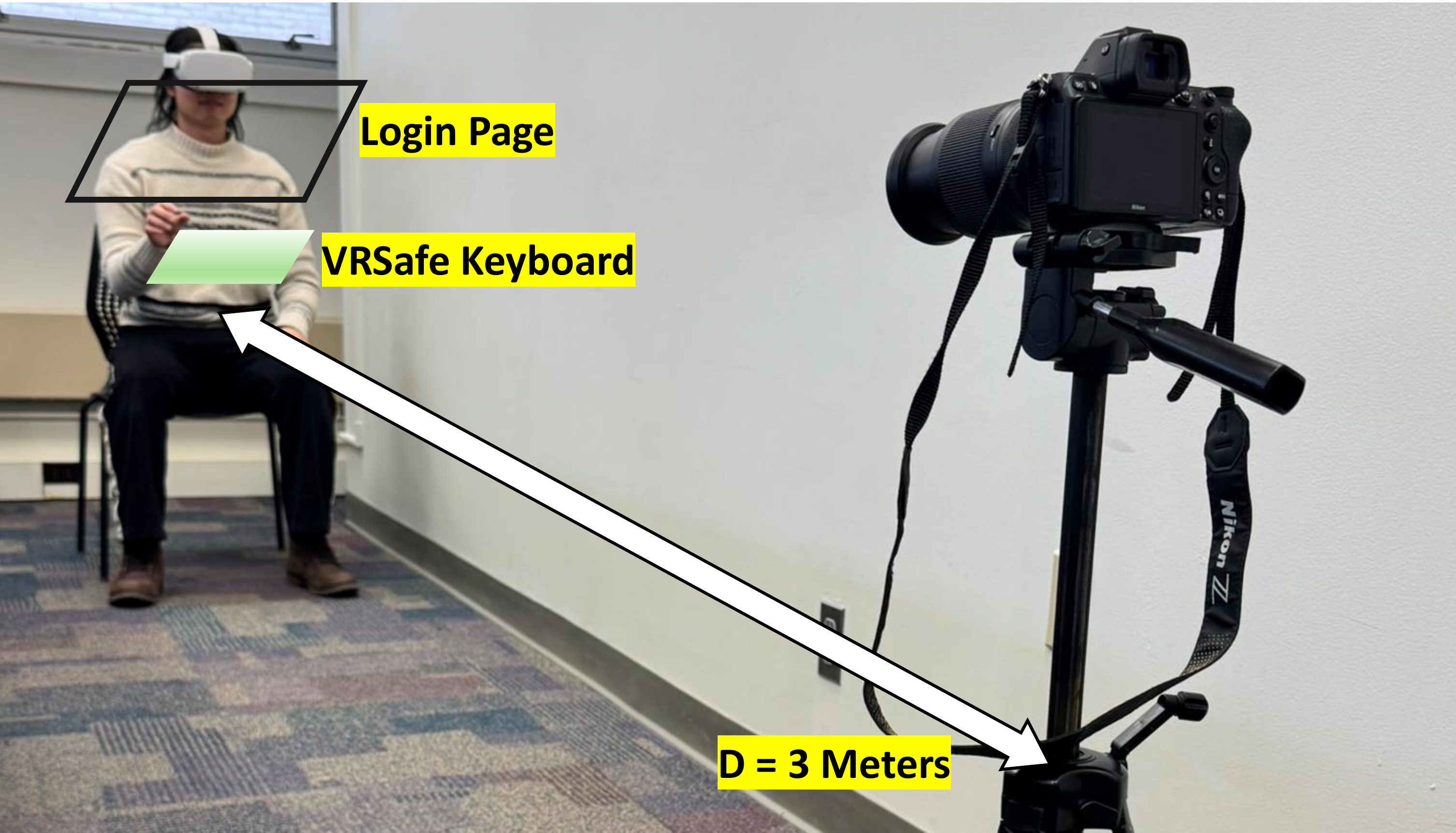}
  \caption{Camera setup for video-based keystroke inference simulation.}
  \Description{Camera position in user study.}
  \label{fig:Camera}
\end{figure}

\begin{table}[!ht]
\centering
\small
\setlength{\tabcolsep}{3pt}
\caption{Reliability and validity test result (Cronbach's Alpha and AVE)}
\label{tab:Cronbach}
\begin{tabular}{p{1.8cm}p{4.5cm}cc}
\hline
\multicolumn{1}{c}{\textbf{Variable}} & 
\multicolumn{1}{c}{\textbf{Measurement Indicator}} & 
\textbf{AVE} & 
\textbf{$\alpha$} \\
\hline
\multirow{4}{1.8cm}{\centering Perceived Ease of Use (PEOU)}
    & My interaction with VR Keyboard is clear and understandable (PEOU1) & \multirow{4}{*}{0.641} & \multirow{4}{*}{0.865} \\
    & Learning to operate the VR Keyboard is easy for me (PEOU2) & & \\
    & The VR Keyboard is user-friendly and requires little effort to understand (PEOU3) & & \\
    & I find it easy to get the VR Keyboard to do what I want it to do (PEOU4) & & \\
\hline
\multirow{3}{1.8cm}{\centering Perceived Usefulness (PU)}
    & Using the VR Keyboard would make me feel safer when entering password (PU1) & \multirow{3}{*}{0.602} & \multirow{3}{*}{0.786} \\
    & Using the VR keyboard helps me feel more confident that my passwords are protected (PU2) & & \\
    & The VR keyboard provides a safer way to enter my passwords compared to traditional methods (PU3) & & \\
\hline
\multirow{2}{1.8cm}{\centering Behavioral Intention To Use (BI)}
    & I expect my use of the VR Keyboard to continue in the future (BI1) & \multirow{2}{*}{0.736} & \multirow{2}{*}{0.852} \\
    & I would recommend this VR Keyboard to others (BI2) & & \\
\hline
\multicolumn{4}{l}{\footnotesize AVE: Average Variance Extracted.} \\
\multicolumn{4}{l}{\footnotesize $\alpha$: Cronbach's Alpha.} \\
\end{tabular}
\end{table}

\onecolumn

\begin{table}[h]
\caption{Survey questions used in the user study.}
\label{tab:vr_keyboard_survey}

\begin{tabular}{|p{0.65\textwidth}|c|c|c|c|c|}
\hline
\textbf{Statement} & \textbf{1} & \textbf{2} & \textbf{3} & \textbf{4} & \textbf{5} \\
\hline
\multicolumn{6}{|l|}{\textbf{Technology Acceptance Model (TAM)}} \\
\hline
Learning to operate the VR Keyboard is easy for me. & & & & & \\
\hline
The VR Keyboard is user-friendly and easy to understand. & & & & & \\
\hline
I find it easy to get the VR Keyboard to do what I want. & & & & & \\
\hline
Using the VR Keyboard makes me feel safer when entering passwords. & & & & & \\
\hline
I am confident that my passwords are protected when using the VR Keyboard. & & & & & \\
\hline
The VR Keyboard provides a safer password-entry method than traditional approaches. & & & & & \\
\hline
I expect to continue using the VR Keyboard in the future. & & & & & \\
\hline
I would recommend the VR Keyboard to others. & & & & & \\
\hline
\multicolumn{6}{|l|}{\textbf{System Usability Scale (SUS)}} \\
\hline
I think I would like to use this VR Keyboard frequently. & & & & & \\
\hline
I found the VR Keyboard unnecessarily complex. & & & & & \\
\hline
I thought the VR Keyboard was easy to use. & & & & & \\
\hline
I would need technical support to use the VR Keyboard. & & & & & \\
\hline
The functions of the VR Keyboard are well integrated. & & & & & \\
\hline
There is too much inconsistency in the VR Keyboard. & & & & & \\
\hline
Most people would learn to use this VR Keyboard very quickly. & & & & & \\
\hline
I found the VR Keyboard cumbersome to use. & & & & & \\
\hline
I felt confident using the VR Keyboard. & & & & & \\
\hline
I needed to learn many things before I could use the VR Keyboard effectively. & & & & & \\
\hline
\multicolumn{6}{|l|}{\textbf{Open-ended Questions}} \\
\hline
For different accounts, do you have different security needs? What is your strategy?
& \multicolumn{5}{c|}{} \\
\hline
For the proposed keyboard design, higher security often requires additional effort. What is your strategy for different password-entry needs?
& \multicolumn{5}{c|}{} \\
\hline
In your opinion, what are the pros and cons of the new design?
& \multicolumn{5}{c|}{} \\
\hline
\end{tabular}
\end{table}

\end{document}